\documentclass[11pt]{article}
\usepackage{epsf,amsmath,amsfonts}

\newcommand{\resection}[1]{\setcounter{equation}{0}\section{#1}}
\newcommand{\bc}[6]{{\,C^{(\!#1\hspace{-0.3pt}#2\hspace{-0.3pt
     }#3\!)#6}_{\;#4#5}\,}}

\newcommand{\eps}{\varepsilon}
\newcommand{\lam}{\lambda}

\newcommand{\tphi}{{\tilde{\phi}}}
\newcommand{\hmu}{{\hat{\mu}}}

\newcommand{\One}{{\hbox{{\rm 1{\hbox to 1.5pt{\hss\rm1}}}}}}

\begin{document}
\begin{center}
{\Large {\bf On Perturbations of Unitary Minimal Models by Boundary
    Condition Changing Operators}} \\[5pt]
\end{center}
\vskip 1.8cm
\centerline{K.~Graham\footnote{e-mail: {\tt kgraham@mth.kcl.ac.uk}}}
\vskip 0.6cm
\centerline{\today }
\centerline{\sl Mathematics Department, }
\centerline{\sl King's College London, Strand, London WC2R 2LS, U.K.}
\vskip 0.9cm
\begin{abstract}
\vskip0.15cm \noindent In this note we consider boundary perturbations in the
A-Series unitary minimal models by $\phi_{r,r+2}$ fields on superpositions of
boundaries.  In particular, we consider perturbations by boundary condition
changing operators.   Within conformal perturbation theory we
explicitly map out the space of perturbative renormalisation group flows for the example $\phi_{13}$ and find that this sheds light on more general $\phi_{r,r+2}$ perturbations.   Finally, we find a simple diagrammatic representation for the space of flows from a single Cardy boundary condition.
\end{abstract}
\setcounter{footnote}{0}
\def\thefootnote{\fnsymbol{footnote}}

\resection{Introduction}
\label{sec:intro}

The study of boundary perturbations in rational conformal field theory has found applications in many areas of string theory, condensed matter physics and statistical mechanics.  Tachyon condensation and the Kondo problem being but two.  Previous investigations of boundary perturbations have concentrated on systems with integrability where there are many powerful techniques available.  More general studies began with \cite{Aff91,Aff93} and were continued in \cite{Rec00,Aff00,Gra01}.  In these latter works, the integral role of superpositions of boundaries came to the fore and has provided the impetus for the present work. 

In this note we consider boundary perturbations in the unitary A-series Virasoro minimal models by $\phi_{r,r+2}$ fields on a superposition of boundary conditions.  In particular this includes perturbations by boundary condition changing operators.  The paper begins with an introduction to the Virasoro minimal models, their boundaries and their boundary condition changing operators (henceforth called simply ``boundary changing operators'').  Then in section \ref{sec:pert} we look at the general theory of $\phi_{r,r+2}$ perturbations within perturbation theory, illustrating it with two examples: $\phi_{13}$ boundary condition changing perturbations and $\phi_{35}$ perturbations on the boundary condition $(2,p)$, $p>2$.  Here we are led to consider perturbations by irrelevant operators.  In section \ref{sec:fp} we find the fixed points of our example systems before identifying the associated boundary theories in sections \ref{sec:nfromb} and \ref{sec:nfromg}.  There is an intriguing relation between our flows and the lattice model studies of Behrend and Pearce \cite{Beh00}.  The relation takes the form of a simple set of diagrammatic rules that we claim, describe the space of renormalisation group flows for a general Cardy boundary condition.  This is the subject of section \ref{sec:connect}.  We conclude with a discussion and directions for future research.

\resection{Unitary Minimal Models and their Boundaries}
\label{sec:mm}

This section reviews a few facts about the unitary A-series Virasoro minimal model $M_{m,m+1}$, with central charge  $c=1-\tfrac{6}{m(m+1)}$.  The primary fields of the bulk theory are characterised by their conformal weight,
\begin{align}
  h_{r,s} = \frac{1}{4m(m+1)} \left[ (r(m+1)-sm)^2-1 \right] \; ,
  \label{eq:h}
\end{align}
where the two integers $r$ and $s$ are taken from the Kac table, $(r,s) \in K = \{(r',s') : 1 \le r'
  \le m-1,   1 \le s' \le m \}$ modulo the symmetry $(r,s) \to
  (m-r,m+1-s)$.  We define $K' \subset K$ such that only one of each pair $\{ (r,s),(m-r,m+1-s) \}$ is included in $K'$.

All conformal boundary conditions can be built out of so-called ``Cardy''
boundary conditions \cite{Car89} which are defined to have a non-degenerate vacuum sector.  Each Cardy boundary condition is in one-to-one
correspondence with a primary field of the bulk theory and so are also
labeled by the Kac table.  A general boundary condition is then a linear
combination of Cardy boundary conditions with positive integer coefficients.

The spectrum of boundary fields on a general boundary condition
$a=\oplus_{i=1}^{n} a_i$, \, $a_i=(r_i,s_i)$, is succinctly represented by the
cylinder partition function,
\begin{align}
  Z_{a a}(q) = \sum_{i,j} Z_{a_i a_j}(q) = \sum_{\tiny{\begin{array}{c} i,j \\ b \in K' \end{array}}} n_{b a_i}{}^{a_j} \chi_b(q) \; ,
  \label{eq:cylZ}
\end{align}
where $q=e^{-\pi L/R}$, $L,R$ are the circumference and length of the cylinder respectively, $\chi_b(q)$ is the
character of the representation $b$ and $n_{b a_i}{}^{a_j}$ are the fusion numbers of the minimal model (see e.g. \cite{Beh99}),
\begin{align}
  n_{(r,s) (r',s')}{}^{(r'',s'')}= N(m)_{r,r'}{}^{r''} N(m+1)_{s,s'}{}^{s''} + N(m)_{m-r,r'}{}^{r''} N(m+1)_{m+1-s,s'}{}^{s''} \; ,
  \label{eq:littlen}
\end{align}
where,
\begin{align}
   N(m)_{r,r'}{}^{r''} = 
  \begin{cases}
    1 \hspace{10mm} r+r'+r'' \text{ odd},\; |r-r'| < r'' \text{ and } 
      r'' < r+r' < 2 m - r'' \;, \\
    0 \hspace{10mm} \text{ otherwise.}
  \end{cases}
  \label{eq:bign}
\end{align}
are the fusion coefficients for $\widehat{su}(2)_{m-2}$.

One of the lessons to take from \eqref{eq:cylZ} is that when considering
superpositions of boundaries, one must also take into account boundary
changing operators as an integral part of the theory.
For example, consider the infinite strip with boundary conditions $a$ on
the left and $b$ on the right.  Moving to the disc we see the
points at infinity are mapped to a pair of operators inserted on the
boundary which have different boundary conditions on either side of them.
Rotate the system slightly and then map back to the strip.  These insertions
are now at a finite distance along the strip and can be seen to 
act in the Hilbert space of the theory: taking elements from the Hilbert
space of the strip with boundary conditions $a$ and $a$, $H_{aa}$ into $H_{ab}$ say.  These are boundary changing operators.

A quantity central to the following is the boundary entropy $g$ of
Affleck and Ludwig, \cite{Aff91,Aff93}, whose value on a $(r,s)$ boundary condition is
given by,
\begin{align}
  \langle \One \rangle_{(r,s)} = g_{(r,s)} = \left( \frac{8}{m(m+1)}
  \right)^{\tfrac{1}{4}} \frac{\sin{\left( \tfrac{\pi r}{m}\right) }
  \sin{\left( \tfrac{\pi s}{m+1}\right) }}{\sqrt{ \sin{\left(
  \tfrac{\pi }{m}\right) } \sin{\left( \tfrac{\pi }{m+1}\right) }}}
  \label{eq:mmg} \; .
\end{align}
On a superposition, its value is simply the sum of the values of the
component Cardy boundaries.  We will also require the one point function of primary bulk fields on the (unit) disc,
\begin{align}
  \langle \Phi_{r,s}(z,\bar{z}) \rangle_\omega = \frac{A_\omega^{(r,s)}}{|1-z\bar{z}|^{2h_{r,s}}} \; ,
\end{align}
where for a Cardy boundary,
\begin{align}
  A_{(r',s')}^{(r,s)} = (-1)^{(s'+r')(r+s)} \left( \frac{8}{m(m+1)}
  \right)^{\tfrac{1}{4}} \frac{\sin{\left( \tfrac{\pi r r' }{m}\right) }
  \sin{\left( \tfrac{\pi s s' }{m+1}\right) }}{\sqrt{ \sin{\left(
  \tfrac{\pi r }{m}\right) } \sin{\left( \tfrac{\pi s }{m+1}\right) }}}
  \label{eq:1ptfn} \; ,
\end{align}
while for a superposition, one again sums the contributions of each component boundary.

Finally we set our conventions for boundary fields and for taking
the operator product expansion (OPE).  Fields will be defined on a disc of
circumference $L$, parameterised by $x \in [0,L)$.  We label our boundary
fields of weight $h_i$ by $\phi_i^{ab}(x)$: such a field interpolates between
boundary conditions $a$ in the region $>x$ and $b$ in the region $<x$.  The
OPE of two boundary fields is given by,
\begin{align}
  x>y , \hspace{10mm} \phi_i^{ab}(x) \phi_j^{cd}(y) \sim \sum_k
  (x-y)^{-h_i-h_j+h_k} \delta^{bc} \bc abdijk \phi_k^{ad}(\tfrac 12
  (x+y)) + \text{desc.} \; ,
\end{align}
where ``desc.'' denotes Virasoro descendants of the primary boundary
fields and the structure constants are known from the work of Runkel \cite{Run99}.

\subsection{The $c \to 1$ limit of Boundary Minimal Models}
\label{sec:mm:cto1}

We are interested in the boundary minimal models as $c \to 1$, $m \to \infty$ and so collect here a couple of important results about the limit.  From the partition function \eqref{eq:cylZ} we see that the spectrum of a single Cardy
boundary $(r,s)$ is,
\begin{align}
  \{(r',s') \,+ \text{desc.}\, : \,\, 0<r'< 2r \, , \,\, 0<s'<2s \, , \,\, r',s' \text{
  odd} \} \; ,
\end{align}
while the spectrum of boundary changing operators between $(r_1,s_1)$
and $(r_2,s_2)$ is
\begin{align}
  \left\{ (r',s')\,+ \text{desc.}\, : \,\, 
    \begin{array}{l} |r_1-r_2|<r'< r_1+r_2 \, , \,\, r_1+r_2+r' \, \text{odd} , \\ |s_1-s_2|<s'< s_1+s_2 \, , \,\, s_1+s_2+s' \, \text{odd}
    \end{array}
  \right\} \; ,
\end{align}
where the weights are given by
\begin{align}
  h_{r,s} = \frac{(r-s)^2}{4} + \frac{r^2-s^2}{4(m+1)} +
  \frac{r^2-1}{4(m+1)^2} + \ldots \label{eq:cto1h} \; .
\end{align}
For a general boundary condition there are a finite number of primary
fields and their weights concentrate around the integers and the integers plus
$\tfrac 14$.  From \eqref{eq:cto1h}, the relevant fields with $h \le 1$ fall
into three sets: $\phi_{r,r}$ with $h \sim 0$, $\phi_{r,r+2}$ with $h \sim 1$
(these two sets can exist on a single boundary or as boundary changing
operators)  and $\phi_{r,r \pm 1}$ with $h \sim \tfrac 14$ (which only exist
as boundary changing operators).  As well as these, we will also be interested in the almost relevant fields $\phi_{r+2,r}$ with $h \sim 1$.  We will see in section \ref{sec:dis} that both $\phi_{r,r+2}$ and $\phi_{r+2,r}$ are important for perturbation theory in
$1/(m+1)$. Perturbations by the other fields must be studied by other methods,
see for example \cite{Gra01}.

\subsection{The results of Recknagel {\em et.~al.}}
\label{sec:rec}

In \cite{Rec00}, Recknagel, Roggenkamp and Schomerus (RRS) studied perturbations
of a single Cardy boundary condition by the field $\phi_{13}$ in the $c \to 1$, $m \to \infty$
limit using perturbation theory in $1/(m+1)$.  After choosing a renormalisation scheme in which the
boundary theory had a non-trivial perturbative fixed point, the authors
evaluated the boundary entropy at this new fixed point. This they compared
term by term in $1/(m+1)$ with \eqref{eq:mmg} and so derived a set of integer
equations that allowed them to conjecture the end points of a generic
$\phi_{13}$ flow.   The result,
\begin{align}
  (r,s) \to \oplus_{i=1}^{\min{\{ r,s \}}} (r+s+1-2i,1) \; .
  \label{eq:rrs}
\end{align}

In this paper we will extend these studies to general
$\phi_{r,r+2}$ perturbations and in particular to $\phi_{13}$ boundary changing
operators.  In fact we will find that the latter will have something
to say about the former.

\resection{Boundary Perturbations}
\label{sec:pert}

Here we study perturbations of a boundary conformal field theory on a
disc of circumference $L$ with boundary condition
$\omega=\oplus_{a=1}^n \omega_a$, a superposition of Cardy boundary
conditions.  The perturbed theory is defined by the prescription,
\begin{gather}
  \langle \psi _1 ... \psi_n \rangle _{\omega ; \lambda_\phi }
  =  \langle P \psi _1 ... \psi_n e^{\delta S} \rangle_\omega
  \; , \notag \\
  \delta S = \int_{\partial M} \! dx
  \left[
    \sum_{\phi,a,b} \lambda_\phi^{ab} \eps ^{-y_\phi} \phi^{ab} (x)
    + \sum_a \alpha_a \eps^{-1} \One_a(x)
  \right] \; ,
  \label{eq:presc}
\end{gather}
where P denotes path
ordering, the $\psi_i = \psi_i(z_i,\bar{z}_i)$ are renormalised fields, $y_\phi = 1-h_\phi$, $h_\phi$ is the conformal weight of $\phi$ and
we have introduced a factor of $\eps ^{-y_\psi}$ to make the couplings
dimensionless. We have also separated the identity field of each component
boundary $\omega_a$ (denoted by $\phi_{11}^{aa}=\One_a$ with coupling $\alpha_a$) from the
general sum of perturbing fields. The identity fields will play a specific
role in what is to follow so we identify them with a special
notation.  To preserve unitarity, the perturbing operator is chosen to
be Hermitean thus for the boundary changing couplings we have
$\lam^{ab}=x+iy$, $\lam^{ba}=x-iy$ for real variables $x$ and $y$.  
In
general, \eqref{eq:presc} is UV divergent and must be regularised.  We choose
to restrict integrals of the non-identity fields to the region $ | x_i-x_j
|>\eps$ where $\eps$ acts as our regulator.

We will evaluate \eqref{eq:presc} in perturbation theory as a power series in
$1/(m+1)$ in a renormalisation scheme with non-trivial renormalisation group fixed points such that $\lam=O(1/(m+1))$.  It turns out that such a scheme is defined by
the requirement that\footnote{
  The vertical bar in \eqref{eq:betascheme} denotes ``evaluated at
  constant \ldots ''.  Also $\partial_\eps = \tfrac{\partial}{\partial \eps}$ and later we will use $d_L = \tfrac{d}{dL}$. },
\begin{align}
  \lim_{\eps \to 0} \left. \eps \partial_\eps P e^{\delta S} \right|_{\lam_R} = 0 \; ,
  \label{eq:betascheme}
\end{align}
with the renormalisation conditions
\begin{align}
  \left. \lam_{bare}=\lam_R \right|_{\eps=L} \; ,
  \label{eq:rencon}
\end{align}
where $\lam_{bare} $ is the bare coupling $\lam$ of \eqref{eq:presc}.

The first task is to calculate the $\beta$-functions in this scheme.
To do this we use a trick advocated in \cite{Car88},  whereby we use
the relation $\beta_\lam \equiv L d_L
\lam_R = (\eps \partial_\eps
\lam_{bare}|_{\lam_R})|_{\lam_{bare}=\lam_R}$  (which follows from
\eqref{eq:rencon} and the fact that $\beta_\lam$ is purely a
function of $\lam_R$) and consider \eqref{eq:betascheme} expanded as a power series in the couplings of the non-identity fields.  The first term in this expansion contains only identity fields,
\begin{align}
  0= P \left[ e^{\sum_a \alpha_a \eps^{-1} \int dx \One_a(x) } \sum_a ( \eps \partial_\eps \alpha_a - \alpha_a ) \eps^{-1} \int dx \One_a(x) \right] \; ,
\end{align}
which requires $\beta_{\alpha_a} = \alpha_a$.  Using this result, the first order contribution to \eqref{eq:betascheme} becomes,
\begin{align}
  0= (\eps \partial_\eps \lam_k - y_k \lam_k )
  \eps^{-y_k} \int_0^L \! dx \; e^{\alpha_a \eps^{-1} x \One_a }
  \phi_k^{ac}(x)
  e^{\alpha_c \eps^{-1} (L-x) \One_c } \; . \label{eq:betafirstorder}
\end{align}
Hence $\beta_{\lam_k} = y_k \lam_k$,.  Moving second order we have,
\begin{align}  
\eps \partial_\eps &\left[\phantom{\int_0^L} \hspace{-3ex} \lam_i \lam_j \eps^{-y_i-y_j} \right. \notag \\
  & \times \left. \int_0^L \! dx_1 dx_2 \; \theta( x_2-x_1-\eps ) \,
  e^{\alpha_a \eps^{-1} x_1 \One_a }
  \phi_i^{ab}(x_1) e^{\alpha_b \eps^{-1} (x_2-x_1) \One_b }
  \phi_j^{bc}(x_2) e^{\alpha_c \eps^{-1} (L-x_2) \One_c } \right] \notag \\
  =& - \lam_i \lam_j \eps^{-y_i-y_j}
  \int_0^{L} \! dx \;
  e^{\alpha_a \eps^{-1} x \One_a }
  \phi_i^{ab}(x) e^{\alpha_b }
  \phi_j^{bc}(x+\eps) e^{\alpha_c \eps^{-1} (L-x-\eps) \One_c } \\
  =& - \lam_i \lam_j \eps^{-y_i-y_j} \notag \\
  & \times \int_0^{L} \! dx \;
  e^{\alpha_a \eps^{-1} (x+\tfrac 12 \eps) \One_a }
  \sum_k \bc abcijk e^{\alpha_b - \tfrac 12 (\alpha_a+\alpha_c) } \eps^{y_i+y_j-y_k}
  \phi_k^{ac}(x+\tfrac 12 \eps)
  e^{\alpha_c \eps^{-1} (L-(x+\tfrac 12 \eps)) \One_c } \\
    =& - \lam_i \lam_j \sum_k \eps^{-y_k} \bc abcijk e^{\alpha_b - \tfrac 12 (\alpha_a+\alpha_c) }
  \int_{-\tfrac 12 \eps}^{L-\tfrac 12 \eps} \! dx \;
  e^{\alpha_a \eps^{-1} x \One_a }
  \phi_k^{ac}(x)
  e^{\alpha_c \eps^{-1} (L-x) \One_c } \; , \label{eq:betasecondorder}
\end{align}
where $\theta(x)$ is the usual step function and we have used the lower order results $\eps
\partial_\eps ( \lam \eps^{-y}) = 0$ and  $\eps \partial_\eps ( \alpha
\eps^{-1}) = 0$. From \eqref{eq:betasecondorder} one can read off the second order correction to the $\beta$-functions,
\begin{align}
  \beta_{\lam_i^{ac}} &= y_i \lam_i^{ac} + \sum_{j,k}  \bc abcjki
  \lam_j^{ab} \lam_k^{bc}  e^{\alpha_b - \tfrac 12
  (\alpha_a+\alpha_c) } \label{eq:betafn} \; , \\
  \beta_{\alpha_a} &= \alpha_a + \sum_{j,k}  \bc abajk\One
  \lam_j^{ab} \lam_k^{ba}  e^{\alpha_b - \alpha_a }
  \label{eq:betafnid} \; ,
\end{align}
where we have added the boundary indices to the coupling
constants.  We should also calculate the third order corrections to the
identity field, however they only play a passive role in the
calculations to follow so we do not write
them explicitly.

\subsection{What can we study from \eqref{eq:betafn} and \eqref{eq:betafnid}?}
\label{sec:dis}

We are interested in perturbation theory, hence for the formulae derived so
far to be valid, the couplings must be small\footnote{If we are careful we could allow the identity field couplings to be large since we have treated those non-perturbatively.  However in the examples considered in this paper, the only fixed points that are both non-perturbative in the identity fields and perturbative in the non-identity fields involve setting all the boundary changing operators to zero.  In this case the perturbation by identity fields acts trivially, projecting onto some subset of component boundaries}.  Let us assume all the couplings in the theory are of order $y \equiv \tfrac{2}{m+1}=y_{13}$ and look for fixed points in (\ref{eq:betafn}, \ref{eq:betafnid}).  For our purposes, it is sufficient to consider the non-identity $\beta$-functions to $O(y^2)$ and the identity $\beta$-functions to $O(y^3)$.  We call fields with $y_i \sim y$ (namely $\phi_{r,r+2}$ and $\phi_{r+2,r}$) ``perturbative'' while all other fields are ``non-perturbative''.  Assuming the structure constants in (\ref{eq:betafn}, \ref{eq:betafnid}) are of $O(1)$ as $m \to \infty$, the fixed points of the non-perturbative fields are all $O(y^2)$.  Consequently we see that to $O(y^2)$, the non-linear terms in the $\beta$-functions contain only perturbative fields and in particular, we can neglect the exponentials of identity couplings.  Furthermore, the $O(y^3)$ corrections to the identity field $\beta$-functions also contain only perturbative fields.  This is because conformal invariance of the two point function requires $\bc abcjk\One \ne 0 \implies y_j=y_k$. Finally we note that the fixed points of the system (\ref{eq:betafn}, \ref{eq:betafnid}) are determined entirely from the $\beta$-functions of the perturbative fields. \\

Which fields should we include in the sums of (\ref{eq:betafn}, \ref{eq:betafnid})?  Naively one would include only relevant fields, but this would be wrong.  We are considering perturbation theory in $y=2/(m+1)$ so any factors of $\eps^{-y_k}$ for $y_k \sim \pm y$ found in the perturbative expansion should also be expanded $\eps^{-y_k}=1-y_k \ln \eps + \ldots$ and the logarithms subtracted order by order.  This prescription introduces extra divergences due to the fusing of relevant fields to perturbative irrelevant fields.  With this in mind, the sums of (\ref{eq:betafn}, \ref{eq:betafnid}) should include both the relevant and the perturbative irrelevant fields.

One can see this another way.  We are expanding in $1/(m+1)$ about the $c=1$ boundary theory.  At $c=1$ all the perturbative fields are marginal and generate each other as counterterms and so must be included in the $\beta$-functions.  The fact that some of these fields turn out to be irrelevant through higher order corrections is unimportant, they still have a non-trivial effect on the dynamics of the renormalisation group flow. \\

We end this section by emphasizing that there are a finite number of fields in the $\beta$-functions (\ref{eq:betafn}, \ref{eq:betafnid}) and so one can use them to study perturbations by all perturbative fields.  This is in contrast to the bulk studies of Zamolodchikov \cite{Zam87} where as $1/(m+1)$ gets small, the number of relevant bulk fields gets very large and in all but a few cases, the perturbation theory becomes unwieldy.

\subsection{Two Examples, Finding the Fixed Points}
\label{sec:fp}

\subsubsection{Example 1 : Boundary Changing Perturbations}
\label{sec:fpex1}

As our first example, we will consider $\phi_{13}$ perturbations on the boundary condition
$\omega=\oplus_{a=1}^n \omega_a = \oplus_{a=1}^n (r,s+2a-2)$.  The spectrum of $\phi_{13}$ fields consists of,
\begin{itemize}
  \item Boundary changing operators $\phi_{13}^{a,a+1}=\psi_a$ acting between
    the boundaries $\omega_a$ and $\omega_{a+1}$, we denote their coupling by
    $\mu_a$.
  \item Their conjugate fields, $\phi_{13}^{a+1,a}=\psi_a^\dagger$, whose coupling we
    denote by $\mu_a^\dagger$.
  \item An ordinary $\phi_{13}^{a,a}=\phi_a$ field on each boundary with
    coupling $\lam_a$.
\end{itemize}
Couplings to all other fields (except the identity fields) can be consistently set to zero.  Following the
discussion of section \ref{sec:dis}, to find the fixed points we need only consider the
$\beta$-functions of the perturbative fields,
\begin{align}
  \beta_{\lam_a} &= y \lam_a + \bc aaa... \lam_a^2
    + \bc aba...  \mu_a \mu_a^\dagger
    + \bc aca... \mu_c \mu_c^\dagger \; , \label{eq:betlam} \\
  \beta_{\mu_a} &= ( y + \bc aab... \lam_a
    + \bc abb... \lam_b ) \mu_a \; , \label{eq:betmu} \\
  \beta_{\mu_a^\dagger} &= ( y + \bc baa... \lam_a
    + \bc bba... \lam_b ) \mu_a^\dagger \; , \label{eq:betmud} 
\end{align}
where the a dot denotes $\phi_{13}$, $b=a+1$ and $c=a-1$, it is also understood that
$\lam_0=\mu_0=\mu^\dagger_0=\lam_{n+1}=\mu_n=\mu_n^\dagger=0$.  When $s=1$ we also need to account for the fact that there is no $\phi_{13}$ field on the $(1,1)$ boundary.

For the A Series minimal models $\bc baa... = \bc aab...$ and $\bc bba... = \bc abb...
$ so we combine \eqref{eq:betmu} and \eqref{eq:betmud} into a single
equation for $\hmu = (\mu_a \mu_a^\dagger)^{\tfrac 12}$,
\begin{align}
  \beta_{\hmu_a} = ( y + \bc aab... \lam_a
    + \bc abb... \lam_b ) \hmu_a \; .
 \label{eq:betmu2}
\end{align}
We now wish to study the fixed points of (\ref{eq:betlam},\ref{eq:betmu2}).
Without loss of generality we can assume all the couplings to boundary
changing operators are non-zero (for if one were zero, between $\omega_b$ and
$\omega_{b+1}$ say, then the equations decompose into two independent sets,
$\oplus_{a=1}^{b} \omega_a$ and $\oplus_{a=b+1}^{n} \omega_a$).   Hence
$\beta_{\hmu_a}=0$ implies,
\begin{align}
  0 = y + \bc aab... \lam_a + \bc abb... \lam_b ,
  \hspace{10mm} b=a+1 , \hspace{10mm} a=1,\ldots , n-1 \; ,
\end{align}
and we can find all the $\lam_a$ in terms of a particular lambda, $\lam$ say.
Substituting these expressions into \eqref{eq:betlam} for $a=1$ gives
$\hmu_1$ in terms of $\lam$, then by iteration we find all $\hmu$ using
\eqref{eq:betlam} for $a=2,\ldots,n-1$. Finally we are left with
\eqref{eq:betlam} for $\lam_n$, this is a quadratic in $\lam$ which has two solutions.  The structure constants are known from
\cite{Run99} and those relevant to this calculation are collected in appendix
\ref{sec:sc13}.  Inserting them, we find that the boundary changing couplings are the same for each solution,
\begin{align}
   \mu_{a}^* \mu_{a}^\dagger &=\frac{ y^2}{4} \frac{(s+n+a-1)(s+a-1)(n-a)a}{(s+2a-1)
    (s+2a-2)} , \hspace{5mm} a=1, \ldots ,n-1 \; ,
\end{align}
while the ordinary fields are given by,
\begin{align}
  \lam_{a}^{*(1)} &=- \frac{y}{2 \sqrt{2}} 
  \frac{ 2a^2-4a+2as-3s+s^2+ns-n +2 }{\sqrt{s+2a-3} \sqrt{s+2a-1}}, \hspace{5mm}
  a=1, \ldots ,n \; , \\
  \lam_{a}^{*(2)} &= - \frac{y}{2 \sqrt{2}} \frac{ 2 a^2 -4a + 2as -s - sn
  +n+1 }{\sqrt{s+2a-3} \sqrt{s+2a-1}}, \hspace{5mm}
  a=1, \ldots ,n \; . \;
\end{align}
These solutions are also valid in the case $s=1$, in which case the two
solutions $\lam^{*(1)}$ and $\lam^{*(2)}$ become identified and $\lam_1^*$
vanishes because there is no $\phi_{13}$ field on the $(1,1)$ boundary.

A point to note is that we were only able to
solve for $\hmu_a=\mu_a \mu_a^{\dagger}$ which is invariant under
$\mu_a \to e^{i \theta}
\mu_a$, $\mu_a^{\dagger} \to e^{-i \theta} \mu_a^{\dagger}$.
This means that we have in fact found a continuum of solutions with dimension equal
to the number of non-zero boundary changing operators, i.e.  the fixed points
contain marginal deformations.  Although we expect this marginality will be lifted at the next order in perturbation theory (there are no marginal boundary operators in any
minimal model), we will take a moment to study the moduli more carefully.  In
a perturbation involving only one pair of boundary changing operators between
boundaries $a$ and $b$, consider the small change $\mu \to \mu + i \theta \mu$,
$\mu^\dagger \to \mu^\dagger - i \theta \mu^\dagger$.  At the fixed point, this
corresponds to adding the perturbation,
\begin{align}
  \delta S = i \theta \; \eps^{-y} \int_{\partial M} \! dx \left[ \mu \psi -
  \mu^\dagger \psi^\dagger \right] \; .
\end{align}
So we are interested in the nature of the operator $\mu \psi - \mu^\dagger
\psi^\dagger$ at the fixed point.  To this end we consider the
following object at the fixed point,
\begin{align}
  \tfrac{\partial}{\partial x} \langle \left[ \One_a (x) -
  \One_b(x) \right] X \rangle_{\lam^*,\mu^*} \; ,
\end{align}
where $X$ denotes the insertions away from $x$. On a small neighbourhood
$(x_0,x_1)$ of $x$, away from the insertions, the identity field is given by
the perturbative expansion,
\begin{align}
  \left.\One_a (x) \right|_{\lam^*,\mu^*} &= \One_a (x)
  + \lam_a \eps^{-y} \! \int_{x_0}^{x_1} \! dy \; \phi_a(y)
  + \mu \eps^{-y} \! \int_{x}^{x_1} \! dy \; \psi(y)
  + \mu^\dagger \eps^{-y} \! \int_{x_0}^{x} \! dy \; \psi^\dagger(y) +
  O(\lam^2) \; .
\end{align}
Then using $\partial_x \One_a(x)=0$ we find,
\begin{align}
  \left. \tfrac{\partial}{\partial x} \left( \One_a (x) - \One_b(x)
  \right) \right|_{\lam^*,\mu^*}
  \sim - 2 \eps^{-y} (\mu  \psi(x) - \mu^\dagger \psi^\dagger(x)) \; ,
\end{align}
indicating the marginal perturbations are generated by derivative fields; the level
one descendent of a $h \sim 0$ field.  This is as we would expect, the only marginal fields to this order are the level
one descendants of $\phi_{rr}$ fields.  One does not expect
perturbations by  derivative fields to have any effect on the theory, in the language of
\cite{Car96} they are redundant fields, and so one may think of these
``lines of marginal stability'' as coordinate singularities in the
space of boundary field theories.  Even so, they are useful in
identifying which boundary theory is represented at the fixed point, we
will discuss this further in section \ref{sec:nfromb} \\

We end the section with a summary of the fixed points we have found with the
example of $\omega = (1,1) \oplus (1,3) \oplus (1,5)$.  The fixed points, $\{
\lam_2, \lam_3, \mu_1 \mu_1^\dagger , \mu_2 \mu_2^\dagger \}$, for this system
are,
\begin{itemize}
  \item The origin $\{ 0,0,0,0 \}$.
  \item $2$ points studied by RRS
    $\{ -y,0,0,0 \}$, $\{ 0,-\sqrt{3} y,0,0 \}$
    together with the combination \newline
    $\{ -y,-\sqrt{3} y,0,0 \}$.
  \item $1$ boundary changing fixed point
    $\{ -\tfrac 12 y,-\tfrac{\sqrt{3}}{2} y,y^2,\tfrac{5}{12} y^2 \}$.
  \item $3$ boundary changing fixed points involving just two boundaries,
    $\{ -\tfrac 12 y,0,\tfrac 38 y^2,0 \}$, \newline
    $\{ -\tfrac 54 y,-\tfrac{3\sqrt{3}}{4} y,0,\tfrac{5}{16} y^2 \}$,
    $\{ \tfrac 14 y,-\tfrac{\sqrt{3}}{4} y,0,\tfrac{5}{16} y^2 \}$.
  \item Finally, the $(1,1)\oplus (1,3)$ boundary changing fixed point with a
    RRS flow on the remaining $(1,5)$ boundary,
    $\{ -\tfrac 12 y,-\sqrt{3} y,\tfrac 38 y^2,0 \}$.
\end{itemize}
All in all, $9$ fixed points.  The general picture is clear.  Our task now is
to identify which boundary condition is present at each fixed point.  This
will be the subject of sections \ref{sec:nfromb} and \ref{sec:nfromg}.

\subsubsection{Example 2 : $\phi_{35}$ Perturbations}
\label{sec:fpex2}

Our second example will be perturbation theory on the boundary condition $\omega=(2,p)$.
From the discussion in section \ref{sec:dis} we only need to consider the
$\beta$-functions of the perturbative fields:  $\{ \phi = \phi_{13},\; \tphi =
\phi_{31},\; d_3 = \tfrac{m+1}{2} L_{-1}\phi_{33},\; \psi =
\phi_{35} \}$ with couplings $\{ \lam_\phi, \lam_\tphi {}, \lam_d,
\lam_\psi \}$ respectively.  The normalisation of the derivative
field follows from the choice $\bc ...dd\One = 1$.  Also unitarity
requires that $\lam_d$ be purely imaginary.  Because the descendent field is redundant, we may consistently set its coupling to zero.  However, we choose not to so that parallels with the boundary changing case are easier to see.

The system has the $\beta$-functions,
\begin{align}
  \beta_\phi &= y \lam_\phi + \bc ...{\phi}{\phi}{\phi} \lam_\phi^2
    + \bc ...{d}{d}{\phi} \lam_d^2
    + \bc ...{\psi}{\psi}{\phi} \lam_\psi^2 \; , \label{eq:beta13} \\
  \beta_\tphi &= -y \lam_\tphi + \bc ...{\tphi}{\tphi}{\tphi} \lam_\tphi^2
    + \bc ...{d}{d}{\tphi} \lam_d^2
    +  \bc ...{\psi}{\psi}{\tphi} \lam_\psi^2 \; , \label{eq:beta31} \\
  \beta_d &= 2 \bc ...{\phi}{d}{d} \lam_\phi \lam_d
    + 2 \bc ...{\tphi}{d}{d} \lam_\tphi \lam_d
    + 2 \bc ...{\psi}{d}{d} \lam_\psi \lam_d \; , \label{eq:beta33} \\
  \beta_\psi &= 2y \lam_\psi
    + \bc ...{\psi}{\psi}{\psi} \lam_\psi^2
    + \bc ...{d}{d}{\psi} \lam_d^2 
    + 2 \bc ...{\phi}{\psi}{\psi} \lam_\phi \lam_\psi
    + 2 \bc ...{\tphi}{\psi}{\psi} \lam_\tphi \lam_\psi \; ,
    \label{eq:beta35}
\end{align}
where in this case the dots denote the $(2,p)$ boundary.  The
structure constants are collected in appendix \ref{sec:sc2p}.  Inserting them into (\ref{eq:beta13}-\ref{eq:beta35}) we obtain the following four isolated fixed points,
\begin{align}
  \{  \lam_\phi, \lam_\tphi ,\lam_d, \lam_\psi \} &=
  \begin{cases}
    f_1 = \{ 0,\; \tfrac 12 \sqrt{\tfrac 32} y ,0,0 \} \; , \\
    f_2 = \{ -\tfrac{ (p-2)\sqrt{p^2-1} }{ 4 \sqrt{2} p } y,
      \; \tfrac{(p+1)(p+2)}{4 \sqrt{6} p} y,
      0,
      \; \tfrac{ \sqrt{ (p^2-4)(p^2-1)} }{4 \sqrt{3} p }y \} \; , \\
    f_3 = \{ -\tfrac{ (p+2)\sqrt{p^2-1} }{ 4 \sqrt{2} p }y,
      -\tfrac{(p-1)(p-2)}{4 \sqrt{6} p}y,
      0, -\tfrac{ \sqrt{(p^2-4)(p^2-1)} }{4 \sqrt{3} p }y \} \; , \\
    f_4 = \{ -\tfrac{\sqrt{p^2-1}}{2\sqrt{2}} y,0,0,0 \} \; ,
   \end{cases}
\end{align}
together with two continua due to the derivative perturbation which we
parameterise by $\theta \in [0,2 \pi)$, with $A(\theta)= 1 - \cos{\theta}$,
\begin{align} 
  \{ \lam_\phi, \lam_\tphi ,&\lam_d, \lam_\psi \} = \\
  &\begin{cases}
    f_5 = \{ - A \tfrac{ \sqrt{p^2-1} }{\sqrt{2} p^2} y ,\; 
      A \tfrac{ (p^2-1) }{\sqrt{6} p^2} y, \;
      i \sin{\theta} \tfrac{ \sqrt{p^2-1} }{ 2p } y,\;
      -A \tfrac{ \sqrt{(p^2-1)(p^2-4)} }{2 \sqrt{3}  p^2 } y \} \; , \\
    f_6 = \{ -(p^2-2A ) \tfrac{ \sqrt{p^2-1} }{ 2\sqrt{2} p^2} y ,\;
      \sqrt{ \tfrac 38 }y - A  \tfrac{(p^2-1)}{\sqrt{6} p^2} y  , \;
      i \sin{\theta} \tfrac{ \sqrt{p^2-1} }{ 2p } y,\;
      A \tfrac{ \sqrt{(p^2-1)(p^2-4)} }{2 \sqrt{3} p^2 }y \} \; .
  \end{cases}
\end{align}
The origin is contained as $f_5(\theta=0)$.

\section{A Study of the $\beta$-functions}
\label{sec:nfromb}

Having found the fixed points, we would now like to identify the associated conformal boundary theories and discover which points are connected to which by relevant flows.   
The fact that the first order term in the $\beta$-function is
universal provides a way of studying the nature of a fixed point directly
from the $\beta$-functions.  Let $\lam^*$ be a non-trivial fixed point for the generic set of equations,
\begin{align}
  \beta_{\lam_i} = y_i \lam_i + \sum_{j,k} C_{jk}{}^i \lam_j \lam_k \; .
  \label{eq:cardybeta}
\end{align}
A couple of remarks:  (i)  We imagine \eqref{eq:cardybeta} to contain the $\beta$-functions for all fields - both relevant and irrelevant.  (ii)  The constants $C_{jk}{}^i$ depend on the choice of renormalisation scheme and will in general differ from those in the OPE.

Re-expressing \eqref{eq:cardybeta} about the new fixed point, $\lam \to \lam^* +
\delta \lam$, we find
\begin{align}
  \beta_{\delta \lam_i} =
    \sum_j \left.  \frac{\partial \beta_{\lam_i}}{\partial \lam_j}
    \right|_{\lam^*} \delta \lam_j + O(\delta \lam^2) \; .
\end{align}
Now by diagonalising the first term, we return the $\beta$-function to the
generic form \eqref{eq:cardybeta} and can read off the spectrum of fields.

Looking for fixed points in \eqref{eq:cardybeta} we know only the perturbative
fields have $\lam^*_a \sim y$ while all other fields have $\lam_n^* \sim y^2$,
so to first order in $y$,
\begin{align}
  B_{ij}^{*} \equiv
    \left. \frac{\partial \beta_{\lam_i}}{\partial \lam_j}
    \right|_{\lam^*}
  = y_i \delta_{ij} + \sum_{a} ( C_{ja}{}^{i} +  C_{aj}{}^{i} ) \lam_a^* \; .
\end{align}
An application of Rayleigh-Schrodinger, gives the change in
$y_n$ for the non-perturbative fields along a flow,
\begin{align}
  \delta y_n =  \sum_{a} (  C_{na}{}^{n} +  C_{an}{}^{n} )  \lam_a^* \; .
  \label{eq:changey}
\end{align}

More interesting are the perturbative fields with $y_i \sim 0$.  For
example, to the order we are considering,
\begin{align}
  \dim \ker{B^{*UV}_{a,b}} - \dim \ker{B^{*IR}_{a,b}} = \text{The change in
  the number of Cardy boundaries} \; .
  \label{eq:ker}
\end{align}
To see this note that $\dim \ker B^*_{a,b}$ counts the number of fields with
$y_i = 0$.  Such fields are the level one descendants of the fields
$\phi_{rr}$, $r \ne 1$.  From \eqref{eq:changey} and \eqref{eq:cto1h} we see
that the total number of $\phi_{rr}$ fields (including the identity) remains constant between
perturbative fixed points so any increase in the number of $y_i=0$ fields must
correspond to a decrease in the number of $\phi_{11}$ fields. The number of
$\phi_{11}$ fields on a boundary condition represents the number of Cardy
boundary conditions in the superposition giving \eqref{eq:ker}.

\subsection{Example 1 : Boundary Changing Perturbations}
\label{sec:nfrombex1}

We now apply this theory to the fixed points found in section \ref{sec:fpex1}.
There it was noted that fixed points involving boundary changing operators
actually contain lines of marginal stability.  In the present discussion this
corresponds to zero eigenvalues of $B^*_{i,j}$, one for each pair of non-zero
boundary changing operators.  This observation allows us to deduce the nature
of the fixed points in the special case of $\omega =
\oplus_{a=1}^n (1,s+2a-2)$, for which the only perturbative fields are
$\phi_{13}$.  In this instance, each conjugate pair of non-zero boundary
changing couplings contributes a zero eigenvalue to $B^*_{i,j}$, so increases the difference between the number of boundaries in the
starting and ending superpositions by one. If all the
boundary changing operators are non-zero then the endpoint will be a single
boundary, and its spectrum must contain $n$ $\phi_{rr}$ fields.  This requires the end point boundary to be either $(n,p)$, $(p,n)$, $(m-n,p)$ or $(m-p,n)$ for some $p \ge n$.   To find $p$ and deduce which of the four options is taken we need some more information.  This is provided by the boundary entropy and the one point functions of bulk fields as will be calculated in section \ref{sec:nfromg}.  For now we quote the zeroth order constraints\footnote{The zeroth order constraints follow simply from the fact we are doing perturbation theory in $1/(m+1)$ and do not require any serious computation.}.  In the case of the boundary entropy, we have that in a perturbative flow from $\omega=\oplus_{a=1}^n (r_a,s_a)$ to
$\omega'=(r',s')$ or $\omega'=(m-r',s')$ satisfies,
\begin{align}
  \sum_{a=1}^n r_a s_a = r'_a s'_a \; .
  \label{eq:gconstr}
\end{align}
This implies the $p=s+n-1$.  To narrow things down further we consider the zeroth order constraint coming from the bulk one point functions \eqref{eq:1ptfn}, allowing us to discount $(m-s-n+1,n)$ and $(m-n,s+n-1)$.  Finally, we
diagonalise $B^*_{a,b}$ and quote the results in the following table.  From
the spectrum we predict the endpoint.
{   \refstepcounter{table}\label{tab:13a}
\begin{center} \begin{tabular}{lllll}
  \hline & \\[-10pt] n&  Spectrum of $\lam^{*(1)}$ &Endpoint & Spectrum
  of $\lam^{*(2)}$ &Endpoint \\ \hline & \\[-10pt] 
  $2$& $0,\pm y,-2y$&
    $(s+1,2)$& $0,\pm y,2y$& $(2,s+1)$ \\ 
  $3$& $0,0,\pm y,\pm 2y,-3y$&
    $(s+2,3)$& $0,0,\pm y,\pm 2y, 3y$& $(3,s+2)$\\ 
  $4$& $0,0,0,\pm y,\pm 2y,\pm 3y,-4y$& $(s+3,4)$&
    $0,0,0,\pm y,\pm 2y,\pm 3y, 4y$& $(4,s+3)$ \\  \hline
\end{tabular}\\[+10pt]
  {\em Table \ref{tab:13a} : The Identification of the fixed points of the
    $\oplus_{a=1}^n (1,s+2a-2)$ systems.}
\end{center}   }

The general pattern is clear and will be shown when we calculate the boundary
entropy in section \ref{sec:nfromg}.  For now we notice that any Cardy
boundary condition can be constructed as flow from a superposition of $(1,s)$
boundary conditions.  Thus there is a hope that we can study the perturbations
of a Cardy boundary by looking at the $\beta$-functions of the superposition.  To
this end consider $\omega=(1,p-1)\oplus(1,p+1)$ which contains a boundary
changing flow to $(2,p)$.   The fixed points of flows starting from $\omega$
(by diagonalising as above) can be identified as follows,
{
   \refstepcounter{table}\label{tab:13b}
\begin{center} \begin{tabular}{cccccll}
  \hline & \\[-10pt] &Fixed Point &$\{ \lam_1,\lam_2,\mu \mu^\dagger
  \}$&&&Spectrum &Endpoint \\ \hline & \\[-10pt] $\{$&$ 0,$&$0,$&$0
  $&$\}$ &$y,y,y,y$ &$(1,p-1)\oplus(1,p+1)$ \\ $\{$&$
  -\tfrac{\sqrt{p(p-2)}}{2 \sqrt{2}} y,$&$0,$&$0$&$ \}$  &$
  -y,y,\tfrac 12 py,\tfrac 12 py $  &$(p-1,1)\oplus(1,p+1)$\\ $\{$&$
  0,$&$-\tfrac{\sqrt{p(p+2)}}{2 \sqrt{2}} y,$&$0$&$ \}$  &$ -\tfrac
  12 py,-\tfrac 12 py,-y,y $  &$(1,p-1)\oplus(p+1,1)$\\ $\{$&$
  -\tfrac{\sqrt{p(p-2)}}{2 \sqrt{2}} y,$&$ -\tfrac{\sqrt{p(p+2)}}{2
  \sqrt{2}} y,$&$0 $&$\}$  &$ -y,-y,-y,-y $
  &$(p-1,1)\oplus(p+1,1)$\\ $\{$&$ \tfrac{\sqrt{p-2}}{2 \sqrt{2p}}
  y,$&$ - \tfrac{ \sqrt{p+2}}{2 \sqrt{2p}} y,$&$ \tfrac{p+1}{4p} y^2
  $&$\}$  &$ -y,0,y,2y $ &$(2,p)$\\ $\{$&$- \tfrac{(p+1)
  \sqrt{p-2}}{2 \sqrt{2p}} y,$&$ - \tfrac{(p-1) \sqrt{p+2}}{2
  \sqrt{2p}} y,$&$ \tfrac{p+1}{4p} y^2$&$ \}$ &$ -2y,-y,0,y $
  &$(p,2)$\\ \hline
\end{tabular} \\[+10pt]
  {\em Table \ref{tab:13b} : The Identification of the fixed points of the
    $(1,p-1)\oplus(1,p+1)$ system.}
 \end{center}   }

It is important to connect the fixed points with relevant flows, i.e. show that the fixed points which we have identified can actually be reached by a relevant perturbation.  For the case of $\phi_{13}$ boundary changing perturbations on the boundary condition $\oplus_{a=1}^n (r,s+2a-2)$,  the $\beta$-functions (\ref{eq:betlam}-\ref{eq:betmud}) have the form,
\begin{align}
  \beta_i = y \lam_i + \sum_{j,k} C_{jk}{}^i \lam_j \lam_k \; ,
  \label{eq:betasolved}
\end{align}
for some constant $y$.  Consider a fixed point $\lam^*$, then $\lam_i(L) = \lam_i^* \, x(L)$ is a solution to \eqref{eq:betasolved} if,
\begin{align}
  \beta_i = \lam_i^* L d_L x &= y \, \lam_i^* \, x + \sum_{j,k} C_{jk}{}^i \lam_j^* \lam_k^* \, x^2 
  = y \, \lam_i^* \, x - y \, \lam_i^* \, x^2 \; ,
\end{align}
which is true if,
\begin{align}
  L d_L x &= y \, x (1-x) \; . \label{eq:trueif} 
\end{align}
By solving \eqref{eq:trueif}, it is easy to show that there is a solution flowing from the origin to any non-trivial fixed point of (\ref{eq:betlam}-\ref{eq:betmud}).  Furthermore, we see from appendix \ref{sec:sc13} that the structure constants, and hence the dynamics, are independent of $r$.  We use this freedom to set $r=1$, then among the fixed points of (\ref{eq:betlam}-\ref{eq:betmud}) is the totally stable point $\oplus_{a=1}^n (s+2a-2,1)$ which corresponds to the end of a RRS flow on each component boundary.  Re-diagonalising (\ref{eq:betlam}-\ref{eq:betmud}) at this point we find it to again have the form \eqref{eq:betasolved} but with $y \to -y$.  The argument above then shows that every fixed point of the $\phi_{13}$ system on $\oplus_{a=1}^n (r,s+2a-2)$ contains a flow to some stable superposition of $(a,1)$ boundary conditions possibly with multiplicities.  These multiplicities indicate the appearance of Chan-Paton indices \cite{Pol}.

\subsection{Example 2 : $\phi_{35}$ Perturbations}
\label{sec:nfrombex2}

Using the fixed points found in section \ref{sec:fpex2}, we diagonalise the
$\beta$-functions of the $(2,p)$ boundary and tabulate the result in
table \ref{tab:2p}.
{
  \refstepcounter{table}\label{tab:2p}
  \begin{center} 
  \begin{tabular}{cll}
    \hline & \\[-10pt] Fixed Point &Spectrum &Endpoint \\ \hline &
    \\[-10pt] 
    $f_1$  &$\{ y,y,y,y\}$ &$(1,p-1)\oplus(1,p+1)$ \\ 
    $f_2$  &$\{ -y,y,\tfrac 12 py,\tfrac 12 py \}$ 
           &$(p-1,1)\oplus(1,p+1)$ \\ 
    $f_3$  &$\{ -\tfrac 12 py,-\tfrac 12 py,-y,y \}$
           &$(1,p-1)\oplus(p+1,1)$\\  
    $f_4$  &$\{ -y,-y,-y,-y \}$ &$(p-1,1)\oplus(p+1,1)$\\
    $f_5$  &$\{ -y,0,y,2y \}$ &$(2,p)$\\ 
    $f_6$  &$\{ -2y,-y,0,y \}$ &$(p,2)$\\ \hline
  \end{tabular} \\[+10pt]
  {\em Table \ref{tab:2p} : The Identification of the fixed points of the
    $(2,p)$ system.}
 \end{center}
}

In agreement with table \ref{tab:13b}.  In fact it turns out that the
$\beta$-functions for the $(2,p)$ system (\ref{eq:beta13}-\ref{eq:beta35}) are equal, by a linear redefinition of
the couplings, to the $\beta$-functions of the $(1,p-1)\oplus(1,p+1)$ system (\ref{eq:betlam}-\ref{eq:betmud}): First translate
to the appropriate fixed point, diagonalise and rescale. \\

We are also interested as to which of these fixed points are connected by relevant flows.  We have already seen that the $(1,p-1)\oplus(1,p+1)$ boundary flows to all the others, and that all the fixed points flow to $(p-1,1)\oplus(p+1,1)$.  
To discover which fixed points can be reached from the $(2,p)$ point, we integrate the $\beta$-functions numerically.  In figure \ref{fig:1} we have plotted a series of flows emanating from the relevant directions of $f_5=(2,3)$ (similar pictures exist for other $p$).  It is hoped the reader will get a feel for how the ``out'' surface of the $(2,p)$ boundary is embedded in the 3-dimensional space of couplings and believe that this surface does indeed contain non-trivial fixed points.  Plotted in figure \ref{fig:2} is the projection of these flows onto the $\lam_\phi,\lam_\psi$ plane.
\begin{center}
\begin{tabular}{cc}
  \refstepcounter{figure}
  \label{fig:1}
  \epsfysize 6truecm
  \epsfbox{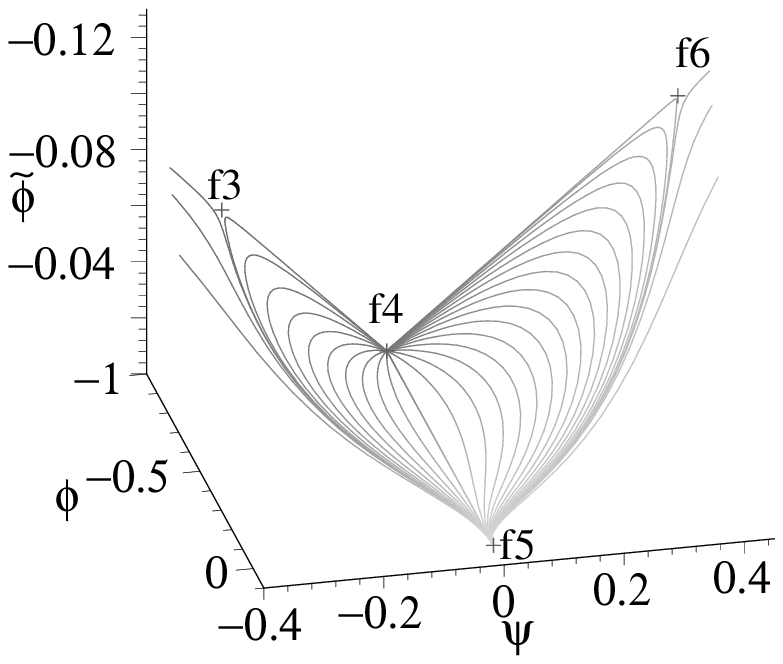}
 & \refstepcounter{figure}
  \label{fig:2}
  \epsfysize 6truecm
  \epsfbox{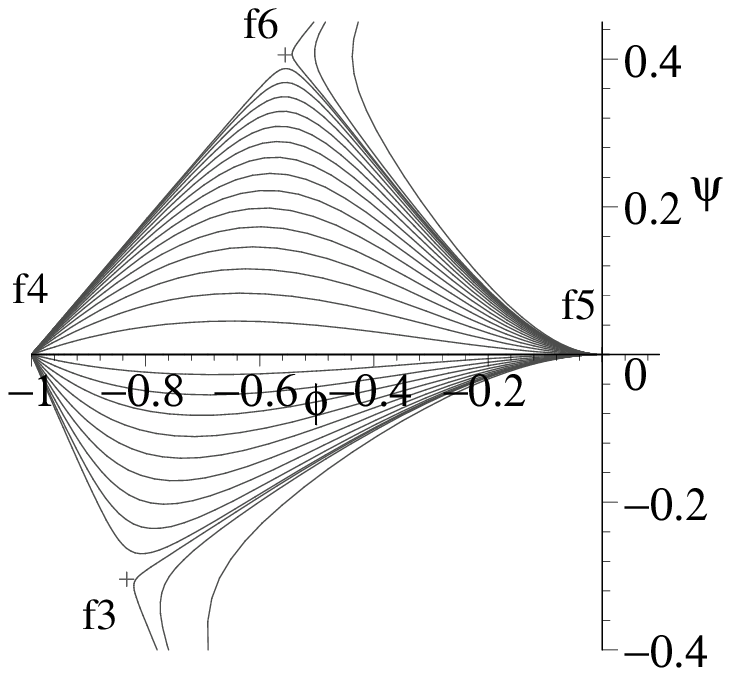}
 \\
 {\em figure \ref{fig:1} : A three dimensional plot} & {\em figure \ref{fig:2} : The projection of} \\
 {\em of a series of flows emanating}& {\em figure  \ref{fig:1} onto the $\phi,\psi$ plane.} \\
 {\em from the $f_5=(2,3)$ fixed point.} & \\ 
\end{tabular}
\end{center} 
 
All in all, we obtain the following picture of flows,
\begin{center}
  \setlength{\unitlength}{1000sp}%
\begingroup\makeatletter\ifx\SetFigFont\undefined%
\gdef\SetFigFont#1#2#3#4#5{%
  \reset@font\fontsize{#1}{#2pt}%
  \fontfamily{#3}\fontseries{#4}\fontshape{#5}%
  \selectfont}%
\fi\endgroup%
\begin{picture}(9622,5878)(1,-8366)
\thicklines
\special{ps: gsave 0 0 0 setrgbcolor}\put(4732,-3134){\vector(-2,-3){545.077}}
\special{ps: grestore}\special{ps: gsave 0 0 0 setrgbcolor}\put(3301,-4861){\vector(-1,-1){900}}
\special{ps: grestore}\special{ps: gsave 0 0 0 setrgbcolor}\put(5101,-4861){\vector( 3,-2){1453.846}}
\special{ps: grestore}\special{ps: gsave 0 0 0 setrgbcolor}\put(2694,-6732){\vector( 1,-3){279.400}}
\special{ps: grestore}\special{ps: gsave 0 0 0 setrgbcolor}\put(9026,-3118){\vector( 1,-2){301.800}}
\special{ps: grestore}\special{ps: gsave 0 0 0 setrgbcolor}\put(8790,-4861){\vector(-1,-2){301.800}}
\special{ps: grestore}\special{ps: gsave 0 0 0 setrgbcolor}\put(7794,-6658){\vector(-1,-3){299}}
\special{ps: grestore}\put(4801,-2836){\makebox(0,0)[lb]{\smash{\SetFigFont{8}{14.4}{\familydefault}{\mddefault}{\updefault}\special{ps: gsave 0 0 0 setrgbcolor}$(1,p-1)\oplus (1,p+1)$\special{ps: grestore}}}}
\put(7426,-4486){\makebox(0,0)[lb]{\smash{\SetFigFont{8}{14.4}{\familydefault}{\mddefault}{\updefault}\special{ps: gsave 0 0 0 setrgbcolor}$(p-1,1)\oplus (1,p+1)$\special{ps: grestore}}}}
\put(7276,-6286){\makebox(0,0)[lb]{\smash{\SetFigFont{8}{14.4}{\familydefault}{\mddefault}{\updefault}\special{ps: gsave 0 0 0 setrgbcolor}$(p,2)$\special{ps: grestore}}}}
\put(  1,-6361){\makebox(0,0)[lb]{\smash{\SetFigFont{8}{14.4}{\familydefault}{\mddefault}{\updefault}\special{ps: gsave 0 0 0 setrgbcolor}$(1,p-1)\oplus (p+1,1)$\special{ps: grestore}}}}
\put(3376,-4561){\makebox(0,0)[lb]{\smash{\SetFigFont{8}{14.4}{\familydefault}{\mddefault}{\updefault}\special{ps: gsave 0 0 0 setrgbcolor}$(2,p)$\special{ps: grestore}}}}
\put(3001,-8236){\makebox(0,0)[lb]{\smash{\SetFigFont{8}{14.4}{\familydefault}{\mddefault}{\updefault}\special{ps: gsave 0 0 0 setrgbcolor}$(p-1,1)\oplus (p+1,1)$\special{ps: grestore}}}}
\end{picture}
\end{center}
where it is true that a flow from $a \to b$ and from $b \to c$ implies that there exists a flow from $a \to c$.  Sadly, numerical studies do not constitute proof that these fixed points can actually be reached.

\resection{Deducing the Nature of the Fixed Points from the Boundary Entropy}
\label{sec:nfromg}

In this section we determine the nature of the fixed points of the boundary changing flows using the boundary entropy following the method of \cite{Rec00}.  The idea is simply to calculate the boundary entropy $g$ in perturbation theory and compare it to the theoretical value \eqref{eq:mmg} term by term in $1/(m+1)$. This gives a set of integer equations which one can then solve.

We consider perturbations of the boundary condition $\omega=\oplus_{a=1}^n
\omega_a = \oplus_{a=1}^n (r,s+2a-2)$ by $\phi_{13}$ fields,
\begin{align}
  \delta S = \int_{\partial M}dx \left[
    \sum_{a=1}^n \left(
      \lam_a \eps ^{-y} \phi_a (x)
      + \alpha_a \eps^{-1} \One_a(x) \right) +
    \sum_{a=1}^{n-1} \left(
      \mu_a \eps^{-y} \psi_a (x)
      + \mu_a^\dagger \eps^{-y} \psi_a^\dagger (x) \right)
  \right] \; ,
  \label{eq:prescbc} \;
\end{align}
with the notation introduced in section \ref{sec:fpex1}.  The quantity
of interest is the boundary entropy,
\begin{align}
  g(\lam,\mu,\alpha) = 
    \sum_{a=1}^n e^{\alpha_a L \eps^{-1} } \langle \One_a \rangle
    &+ \frac{1}{2} \eps^{-2y} \sum_{a=1}^n
      \lam_a^2 e^{ \alpha_a L \eps^{-1}} \!\!\! \int \! dx_1 dx_2
      \langle P \phi_a(x_1) \phi_i(x_2) \rangle  \notag \\
   + \eps^{-2y} \sum_{a=1}^{n-1}
      \mu_a \mu_a^\dagger & \int \! dx_1 dx_2
      \langle P \psi_a(x_1) \psi_a^\dagger(x_2)
        e^{ \eps^{-1} \!\! \int dx \left(
          \alpha_a \One_a(x) + \alpha_{a+1} \One_{a+1}(x) \right) }
      \rangle \notag \\
   + \frac 16 \eps^{-3y} \sum_{a=1}^n
      \lam_a^3  e&^{ \alpha_a L \eps^{-1}} \!\!\! \int \! dx_1 dx_2 dx_3
      \langle P \phi_a(x_1) \phi_a(x_2) \phi_a(x_3) \rangle \notag \\
   + \eps^{-3y} \sum_{a=1}^{n-1} \sum_{b=a,a+1}
      \lam_b \mu_a \mu_a^\dagger \! &\int \! dx_1 dx_2 dx_3 
      \langle P \phi_b(x_1) \psi_a(x_2) \psi_a^\dagger(x_3)
        e^{ \eps^{-1} \!\! \int dx \left(
          \alpha_a \One_a(x) + \alpha_{a+1} \One_{a+1}(x) \right)}
      \rangle \; ,\notag \\
 = g_0(\alpha) + g_2(\lam,\alpha &) + g_2(\mu,\alpha) + g_3(\lam,\alpha)
    + g_3(\mu,\lam,\alpha) \; .
  \label{eq:gpert}
\end{align}
where we have pulled the identity fields out of the correlation functions
where we can.  The divergent integrals are regularised as discussed in section \ref{sec:pert}.

Before continuing, we take a moment to consider the identity fields in
the calculation to follow.  To third order, the $\beta$-functions for
the identity fields \eqref{eq:betafnid} do not contain higher order
identity field corrections.  Because of this it is useful to make a
change of renormalisation scheme to remove the higher order
corrections in the other fields as well, leaving one with a linear
$\beta$-function,
\begin{align}
  \beta_{\alpha'}=\alpha' \; .
\end{align}
To second order, this change is achieved by the transformation\footnote{Here
we have used the fact that all the relevant
  fields have the same weight $y$; the generalisation is
  straightforward as is the inclusion of third order terms.},
\begin{align}
  \alpha'=\alpha + (1-2y)^{-1} \sum_{j,k} C_{jk}{}^{\One} \lam_j \lam_k \; ,
\end{align}
which is non-singular and geometrically corresponds to projecting the
flows onto the $\alpha'$ plane.  Under such a transformation, the renormalisation condition also changes,
\begin{align}
  \alpha_R|_{\eps=L} = \alpha_{bare} \longrightarrow
  \alpha'_R|_{\eps=L} = \alpha_{bare}+(1-2y)^{-1} \sum_{j,k} C_{jk}{}^{\One} \lam_{j,bare} \lam_{k,bare} \; .
\end{align}
To see the effect of the new condition, we solve $\beta_{\alpha'}=Ld_L
\alpha'=\alpha'$,
\begin{align}
 \tfrac{L}{\eps} \alpha_{bare}= \alpha_R' - \tfrac{L}{\eps}
 (1-2y)^{-1} \sum_{j,k} C_{jk}{}^{\One} \lam_{j,bare}
 \lam_{k,bare} \; .
\end{align}
The final term is generally interpreted as the ground state energy
correction.  That is to say, the scheme with a linear
$\beta$-function for the identity fields is the scheme in which the
identity fields are used to absorb the ground state energy corrections
and only these corrections.   Furthermore, at the fixed points
$\alpha_R'=0$.   Because of this we choose simply to ignore both the identity fields
and the ground state
energy corrections knowing that one is taking care of the other.

With that said, the calculation is almost identical to
\cite{Aff93}, from which we take the following results,
\begin{align}
  g_2(\lam_a) &= \tfrac{1}{2} \eps^{-2y} \lam_a^2 \int dx_1 dx_2
    \langle P \phi_a(x_1) \phi_a(x_2) \rangle
    = - \lam_a^2 \pi^2 y \left( \tfrac{\eps}{L} \right)^{-2y}
      \bc aaa..\One \langle \One_a \rangle \; , \\
  g_2(\mu_a) &=  \eps^{-2y} \mu_a \mu_a^\dagger \int dx_1 dx_2
    \langle P \psi_a(x_1) \psi_a^\dagger(x_2) \rangle
    = - \mu_a \mu_a^\dagger \pi^2 y \left( \tfrac{\eps}{L} \right)^{-2y}
      \bc aba..\One \langle \One_a \rangle \; , \\
  g_3(\lam_a) &=  \tfrac{1}{6} \eps^{-3y} \lam_a^3 \int dx_1 dx_2 dx_3
    \langle P \phi_a(x_1) \phi_a(x_2) \phi_a(x_3) \rangle \notag \\
   &= \lam_a^3 \pi^2 \left( \tfrac{\eps}{L} \right)^{-3y}
      \left[ - \tfrac 83 + 2 \left( \tfrac{\eps}{L} \right)^{y} \right]
      \bc aaa... \bc aaa..\One \langle \One_a \rangle \; , \\
  g_3(\lam_a,\mu_b) &=  \eps^{-3y} \lam_a \mu_b \mu_b^\dagger
    \int dx_1 dx_2 dx_3
    \langle P \phi_a(x_1) \psi_b(x_2) \psi_b^\dagger(x_3) \rangle
    \notag \\
   &=3 \lam_a \mu_b \mu_b^\dagger \pi^2 \left( \tfrac{\eps}{L} \right)^{-3y}
      \left[ - \tfrac 83 + 2 \left( \tfrac{\eps}{L} \right)^{y} \right]
      \bc baa... \bc aba..\One \langle \One_a \rangle \; .
\end{align}

We also solve the $\beta$-function equations (see appendix \ref{sec:solbeta}) to express the bare
couplings in terms of the renormalised ones,
\begin{align}
  \left( \tfrac{\eps}{L} \right)^{-y} \lam_a &=  \lam_a^R -
  \frac 1y \left( 1 - \left( \tfrac{\eps}{L} \right)^y
  \right) \left[ \bc aaa... (\lam_a^R)^2 + \bc aba...  \mu_a^R
  \mu_a^{R\dagger}  + \bc aca... \mu_c^R \mu_c^{R\dagger} \right] \; , \notag \\
  \left( \tfrac{\eps}{L} \right)^{-y} \mu_a &=  \mu_a^R -
  \frac 1y \left( 1 - \left( \tfrac{\eps}{L} \right)^y
  \right) \left[\bc aab... \lam_a^R \mu_a^R + \bc
  abb... \lam_b^R \mu_a^R \right] \; , \label{eq:lambare}\\
  \left( \tfrac{\eps}{L} \right)^{-y} \mu^{\dagger}_a &=  \mu_a^{R\dagger} -
  \frac 1y \left( 1 - \left( \tfrac{\eps}{L} \right)^y
  \right) \left[\bc aab... \lam_a^R \mu_a^{R\dagger} + \bc
  abb... \lam_b^R \mu_a^{R\dagger} \right] \; , \notag
\end{align}
where $b=a+1$ and $c=a-1$.   Substituting into \eqref{eq:gpert} and
using $\bc aba..\One \langle \One_a \rangle = \bc bab..\One \langle
\One_b \rangle$ one finds,
\begin{align}
  g(\lam,\mu) &= \sum_{a=1}^n \langle \One_a \rangle
  - \pi^2 y \sum_{a=1}^n (\lam_a^R)^2 \bc aaa..\One \langle \One_a \rangle
  - 2 \pi^2 y \sum_{a=1}^{n-1} \mu_a^R \mu_a^{R\dagger} \bc
  aba..\One \langle \One_a \rangle \notag \\
  &-\frac{2 \pi^{2}}{3} \sum_{a=1}^n (\lam_a^R)^3 \bc aaa... \bc aaa..\One  \langle \One_a
  \rangle  \notag \\
  &- 2 \pi^2 \sum_{a=1}^{n-1} \lam_a^R \mu_a^R \mu_a^{R\dagger} \bc aab... \bc aba..\One
  \langle \One_a \rangle
  - 2 \pi^2 \sum_{a=1}^{n-1} \lam_b^R \mu_a^R \mu_a^{R\dagger} \bc abb... \bc aba..\One
  \langle \One_a \rangle \; . \label{eq:pertg}
\end{align}
In subsequent sections all couplings will be renormalised so we will
drop the $R$.  We end this section by commenting that the expansions \eqref{eq:lambare} are only good for $\lam << \lam^*$.   However when substituted into physical quantities (e.g. \eqref{eq:pertg}), the resulting expressions remain true in the limit $\lam \to \lam^*$.  However to use \eqref{eq:lambare} in this limit, there must exists a $\lam(L)$ such that $\lim_{L \to \infty} \lam(L) = \lam^*$.  Such a $\lam$ was found in section \ref{sec:nfrombex1}. 

\subsection{Discussion of g-theorem}
\label{sec:gthm}

Having calculated $g$, we are almost duty-bound to say a few words
about the $g$-theorem of Affleck and Lugwig \cite{Aff91,Aff93}.  Here we
note that the $\beta$-functions are gradient flows (c.f. \cite{Zam87}),
\begin{align}
  \frac{\partial g}{\partial \lam_i} = - \sum_{j} G_{ij} \beta_j \; ,
\end{align}
where $G$ is a symmetric positive definite matrix with non-zero entries,
\begin{align}
  G_{\lam_a \lam_a} &= 2 \pi^2 \bc aaa..\One \langle \One_a \rangle \; , \\
  G_{\mu_a \mu_a^\dagger} &= 2 \pi^2 \bc aba..\One \langle \One_a
  \rangle = 2 \pi^2 \bc bab..\One \langle \One_b
  \rangle \; , 
\end{align}
and where $b=a+1$.  Thus the $g$-theorem,
\begin{align}
  L \frac{d g}{d L} = - \sum_{i,j} G_{ij} \beta_i \beta_j \le 0 \; ,
\end{align}
follows from unitarity:  The negativity of the contribution from the ordinary fields is
clear.  For the boundary changing operators, unitarity constrains us to write $\mu=x+iy$,
$\mu^\dagger=x-iy$ for real $x$ and $y$ (see section \ref{sec:pert}) and hence
\begin{align}
  \beta_\mu \beta_{\mu^\dagger} = \beta_x^2+\beta_y^2 \; .
\end{align}

\subsection{Evaluation of End-points}
\label{sec:evalend}

All conformal boundary states in the minimal models may be written as a superposition of
Cardy's boundary conditions.   We denote Cardy boundary conditions by Greek
letters, labeled by a superscript.
\begin{alignat}{3}
  \omega &= \oplus_{\ell=1}^n \, \omega^\ell ,\qquad &\alpha &= \oplus_{\ell=1}^{n'} \, \alpha^\ell \; , \\
  \omega^\ell &= (\omega^\ell_1,\omega^\ell_2) = (r,s+2\ell-2),\qquad &\alpha^\ell &= (\alpha^\ell_1,\alpha^\ell_2) \; , 
\end{alignat}
where $\omega$ denotes the initial boundary while $\alpha$ will be the
final boundary.  
The boundary entropy of a superposition of boundary states $\omega$, is given by the sum
\begin{align}
  g^\omega_{sup} = \sum_\ell g_{\omega^\ell} \; .
\end{align}
Using \eqref{eq:mmg}, the object of interest is,
\begin{align}
  \Delta \ln g &=  \ln \left( \frac{g^\alpha_{sup}}{g^\omega_{sup}}
  \right) \notag \\
  &=
  \ln \left( \frac{\sigma_\alpha}{\sigma_\omega} \right)
  + \frac{\pi^2}{6}\left(
  \sum_\ell \frac{ \omega^\ell_1 \omega^\ell_2 }{\sigma_\omega}((\omega^\ell_1)^2 +
  (\omega^\ell_2)^2) -
  \sum_\ell \frac{ \alpha^\ell_1 \alpha^\ell_2}{\sigma_\alpha}((\alpha^\ell_1)^2 +
  (\alpha^\ell_2)^2) \right) (m+1)^{-2} \notag \\
  &+ \frac{\pi^2}{3}\left(
  \sum_\ell \frac{ \omega^\ell_1 \omega^\ell_2 }{\sigma_\omega}(\omega^\ell_1)^2 -
  \sum_\ell \frac{ \alpha^\ell_1 \alpha^\ell_2 }{\sigma_\alpha}((\alpha^\ell_1)^2 \right)
  (m+1)^{-3} + O\left( (m+1)^{-4} \right) \label{eq:zratio} \; ,
\end{align}
where
\begin{align}
  \sigma_{\omega} &= \sum_\ell  \omega^\ell_1 \omega^\ell_2 \; .
\end{align}
From the perturbative calculation, this is given by \eqref{eq:pertg},
\begin{align}
  \Delta \ln g(\lam^*,\mu^*) &= \ln
  \left( \frac{g(\lam^*,\mu^*)}{g^\omega_{sup}} \right) \; . 
  \label{eq:zratiopert}
\end{align}
Without loss of generality, we can assume all the boundary changing
couplings are non-zero.  
Inserting values for the structure constants, the fixed
points found in section \ref{sec:fpex1}, and $g^\omega_{sup}$ one finds the rather simple result,
\begin{align}
   \Delta \ln g(\lam^{*(1)},\mu^*)
  &= - \frac{\pi^2}{3} \frac{(s+n-2)(s+n)}{(m+1)^3} \; , \\
   \Delta \ln g(\lam^{*(2)},\mu^*)
  &= - \frac{\pi^2}{3} \frac{(n-1)(n+1)}{(m+1)^3} \; , 
\end{align}
Equating this with \eqref{eq:zratio} provides the following set of polynomial
equations to be solved over the integers,
\begin{align}
  \sigma_{\omega} &= \sigma_{\alpha} \; , \\
   \sum_\ell \frac{ \omega^\ell_1 \omega^\ell_2
   }{\sigma_{\omega}}((\omega^\ell_1)^2 + (\omega^\ell_2)^2) &=
  \sum_\ell \frac{ \alpha^\ell_1 \alpha^\ell_2}{\sigma_{\alpha}}((\alpha^\ell_1)^2 +
  (\alpha^\ell_2)^2) \; , \\
  \sum_\ell \frac{ \alpha^\ell_1 \alpha^\ell_2 }{\sigma_{\alpha}}((\alpha^\ell_1)^2 -
  \sum_\ell \frac{ \omega^\ell_1 \omega^\ell_2 }{\sigma_{\omega}}(\omega^\ell_1)^2
    &= \begin{cases}
    (s+n-2)(s+n)& \text{for solution 1}, \\
    (n-1)(n+1)& \text{for solution 2}.
  \end{cases} \;
\end{align}

As in \cite{Rec00}, these equations generally have many solutions,
however a generic solution does appear.   For solution 1 we have,
\begin{align}
  \bigoplus_{i=1}^{\min{\{r,s+n-1\}}} (r+s+n-2i,n) \; ,
  \label{eq:ans1}
\end{align}
while for solution 2,
\begin{align}
  \bigoplus_{i=1}^{\min{\{r,n\}}} (r+n+1-2i,s+n-1) \; ,
  \label{eq:ans2}
\end{align}
The calculation of $g$ is ambiguous because of the symmetries $g_{(r,s)}
= g_{(m-r,s)} =  g_{(r,m+1-s)}$.  One may remove this uncertainty by simply calculating the one point functions of bulk fields to zeroth order in perturbation theory.

We complete this section with our example of $\omega = (1,1) \oplus (1,3)
\oplus (1,5)$.  In section \ref{sec:fpex1} we found nine fixed points
for the system, we now identify the boundary theories associated to
these points,
{
   \refstepcounter{table}\label{tab:33}
\begin{center} \begin{tabular}{lll}
 \hline & \\[-10pt] 
 Fixed Point & &Boundary Theory \\
 $\{ \lam_2, \lam_3, \mu_1 \mu_1^\dagger , \mu_2 \mu_2^\dagger \}$& \\
 \hline & \\[-10pt] 
  $\{ 0,0,0,0 \}$ & &$(1,1) \oplus (1,3) \oplus (1,5)$ \\
  $\{ -y,0,0,0 \}$ & & $(1,1) \oplus (3,1) \oplus (1,5)$ \\
  $\{ 0,-\sqrt{3} y,0,0 \}$ & 
    & $(1,1) \oplus (1,3) \oplus (5,1)$ \\
  $\{ -y,-\sqrt{3} y,0,0 \}$ & 
    & $(1,1) \oplus (3,1) \oplus (5,1)$ \\
  $\{ -\tfrac 12 y,-\tfrac{\sqrt{3}}{2} y,y^2,\tfrac{5}{12} y^2 \}$& 
    &$(3,3)$ \\
  $\{ -\tfrac 12 y,0,\tfrac 38 y^2,0 \}$& &$(2,2) \oplus (1,5)$ \\
  $\{ -\tfrac 54 y,-\tfrac{3\sqrt{3}}{4} y,0,\tfrac{5}{16} y^2 \}$&
    &$(1,1) \oplus (2,4)$ \\
  $\{ \tfrac 14 y,-\tfrac{\sqrt{3}}{4} y,0,\tfrac{5}{16} y^2 \}$&
    &$(1,1) \oplus (4,2)$ \\
  $\{ -\tfrac 12 y,-\sqrt{3} y,\tfrac 38 y^2,0 \}$& 
    &$(2,2) \oplus (5,1)$
  \\ \hline
\end{tabular} \\[+10pt]
  {\em Table \ref{tab:33} : The Identification of the fixed points of the
    $(1,1)\oplus (1,3)\oplus (1,5)$ system.}
 \end{center}

\resection{A Connection with Lattice theories}
\label{sec:connect}

With the wealth of fixed points observed in these systems, it would be helpful to find a nice way of organising everything.  In this section, we provide a simple set of rules which we claim, describe the space of flows for an arbitrary Cardy boundary condition.

There is an intriguing connection between the flows found here and the lattice
representation of conformal boundary conditions found in \cite{Beh00} which
we will now explain.  In \cite{Beh00}, each boundary condition was shown to be related to a set of integrable
lattice boundary weights which when taken to the continuum limit, returned the
boundary theory.  As part of their
construction, the authors associate to each Cardy boundary condition in the minimal model $M_{m,m+1}$, a subgraph of (in our case) the Dynkin diagram $A_m$.  It is this subgraph that is of interest to us.

Let us define fused adjacency matrix $F_{ab}^r$ as follows,
\begin{align}
  F_{ab}^r = N(m+1)_{ra}{}^{b} = 
  \begin{cases}
    1 \qquad a+b+r \text{ odd}, |a-b| \le r-1 \text{ and } 
      r+1 \le a+b \le 2 m -r + 1 \; , \\
    0 \qquad \text{ otherwise.}
  \end{cases}
  \label{eq:defnFA}
\end{align}
$F_{ab}^r$ is symmetric in all three indices.

We label the nodes of $A_m$ by $1,2,\ldots,m$ then the subgraph
associated to the Cardy boundary condition $(r,s)$ is the set of
linked nodes,
\begin{align}
  \{ (i,i \pm 1) : F_{is}^r F_{i \pm 1,s}^{r+1} > 0 \} \; .
\end{align}
Distilling this for the case at hand, we see the $(r,s)$ boundary
condition in the limit of large $m$, is represented by the subgraphs,
\begin{center}
  $r \ge s$ \hspace{10mm}\setlength{\unitlength}{1000sp}%
\begingroup\makeatletter\ifx\SetFigFont\undefined%
\gdef\SetFigFont#1#2#3#4#5{%
  \reset@font\fontsize{#1}{#2pt}%
  \fontfamily{#3}\fontseries{#4}\fontshape{#5}%
  \selectfont}%
\fi\endgroup%
\begin{picture}(11298,753)(1651,-8014)
\thicklines
\put(6901,-7861){\circle*{250}}
\put(5401,-7861){\circle*{250}}
\put(2101,-7861){\circle*{50}}
\put(2401,-7861){\circle*{50}}
\put(2701,-7861){\circle*{50}}
\put(4801,-7861){\circle*{50}}
\put(5101,-7861){\circle*{50}}
\put(11101,-7861){\circle*{50}}
\put(11401,-7861){\circle*{50}}
\put(12601,-7861){\circle*{50}}
\put(12901,-7861){\circle*{50}}
\put(11701,-7861){\circle*{50}}
\put(3301,-7861){\circle*{250}}
\put(3001,-7861){\circle*{50}}
\put(1801,-7861){\circle*{250}}
\put(3601,-7861){\circle*{50}}
\put(3901,-7861){\circle*{50}}
\put(9001,-7861){\circle*{250}}
\put(10488,-7874){\circle*{250}}
\put(10801,-7861){\circle*{50}}
\put(12001,-7861){\circle*{250}}
\put(12301,-7861){\circle*{50}}
\special{ps: gsave 0 0 0 setrgbcolor}\put(5401,-7861){\line( 1, 0){2250}}
\special{ps: grestore}\special{ps: gsave 0 0 0 setrgbcolor}\put(8251,-7861){\line( 1, 0){2250}}
\special{ps: grestore}\put(10201,-7561){\makebox(0,0)[lb]{\smash{\SetFigFont{8}{14.4}{\familydefault}{\mddefault}{\updefault}\special{ps: gsave 0 0 0 setrgbcolor}$r+s$\special{ps: grestore}}}}
\put(4651,-7561){\makebox(0,0)[lb]{\smash{\SetFigFont{8}{14.4}{\familydefault}{\mddefault}{\updefault}\special{ps: gsave 0 0 0 setrgbcolor}$r-s+1$\special{ps: grestore}}}}
\put(1651,-7561){\makebox(0,0)[lb]{\smash{\SetFigFont{8}{14.4}{\familydefault}{\mddefault}{\updefault}\special{ps: gsave 0 0 0 setrgbcolor}$1$\special{ps: grestore}}}}
\put(3151,-7561){\makebox(0,0)[lb]{\smash{\SetFigFont{8}{14.4}{\familydefault}{\mddefault}{\updefault}\special{ps: gsave 0 0 0 setrgbcolor}$2$\special{ps: grestore}}}}
\end{picture} \\
\vspace{3mm} 
  $r < s$ \hspace{10mm}\setlength{\unitlength}{1000sp}%
\begingroup\makeatletter\ifx\SetFigFont\undefined%
\gdef\SetFigFont#1#2#3#4#5{%
  \reset@font\fontsize{#1}{#2pt}%
  \fontfamily{#3}\fontseries{#4}\fontshape{#5}%
  \selectfont}%
\fi\endgroup%
\begin{picture}(11298,753)(1651,-8014)
\thicklines
\put(6901,-7861){\circle*{250}}
\put(5401,-7861){\circle*{250}}
\put(2101,-7861){\circle*{50}}
\put(2401,-7861){\circle*{50}}
\put(2701,-7861){\circle*{50}}
\put(4801,-7861){\circle*{50}}
\put(5101,-7861){\circle*{50}}
\put(11101,-7861){\circle*{50}}
\put(11401,-7861){\circle*{50}}
\put(12601,-7861){\circle*{50}}
\put(12901,-7861){\circle*{50}}
\put(11701,-7861){\circle*{50}}
\put(3301,-7861){\circle*{250}}
\put(3001,-7861){\circle*{50}}
\put(1801,-7861){\circle*{250}}
\put(3601,-7861){\circle*{50}}
\put(3901,-7861){\circle*{50}}
\put(9001,-7861){\circle*{250}}
\put(10488,-7874){\circle*{250}}
\put(10801,-7861){\circle*{50}}
\put(12001,-7861){\circle*{250}}
\put(12301,-7861){\circle*{50}}
\special{ps: gsave 0 0 0 setrgbcolor}\put(5401,-7861){\line( 1, 0){2250}}
\special{ps: grestore}\special{ps: gsave 0 0 0 setrgbcolor}\put(8251,-7861){\line( 1, 0){2250}}
\special{ps: grestore}\put(10201,-7561){\makebox(0,0)[lb]{\smash{\SetFigFont{8}{14.4}{\familydefault}{\mddefault}{\updefault}\special{ps: gsave 0 0 0 setrgbcolor}$r+s$\special{ps: grestore}}}}
\put(4851,-7561){\makebox(0,0)[lb]{\smash{\SetFigFont{8}{14.4}{\familydefault}{\mddefault}{\updefault}\special{ps: gsave 0 0 0 setrgbcolor}$s-r$\special{ps: grestore}}}}
\put(1651,-7561){\makebox(0,0)[lb]{\smash{\SetFigFont{8}{14.4}{\familydefault}{\mddefault}{\updefault}\special{ps: gsave 0 0 0 setrgbcolor}$1$\special{ps: grestore}}}}
\put(3151,-7561){\makebox(0,0)[lb]{\smash{\SetFigFont{8}{14.4}{\familydefault}{\mddefault}{\updefault}\special{ps: gsave 0 0 0 setrgbcolor}$2$\special{ps: grestore}}}}
\end{picture} \\
\end{center}
A first indication that these diagrams have relevance to boundary perturbations is the observation that the number of links in the diagram is equal to the number of relevant operators on the boundary (The proof is contained in appendix \ref{sec:lro}).

We now introduce our diagrammatic rules for the flows.  Flows will
correspond to projections onto subgraphs.  Assign an orientation to each
link in the graph whereby the link connected to node
$r+s$ is negative and contiguous
links have opposite orientation, for example for $(3,3)$,
\begin{center}
  \setlength{\unitlength}{1000sp}%
\begingroup\makeatletter\ifx\SetFigFont\undefined%
\gdef\SetFigFont#1#2#3#4#5{%
  \reset@font\fontsize{#1}{#2pt}%
  \fontfamily{#3}\fontseries{#4}\fontshape{#5}%
  \selectfont}%
\fi\endgroup%
\begin{picture}(8580,560)(1661,-8001)
\thicklines
\put(3301,-7861){\circle*{250}}
\put(1801,-7861){\circle*{250}}
\put(4801,-7861){\circle*{250}}
\put(6301,-7861){\circle*{250}}
\put(9301,-7861){\circle*{250}}
\put(9601,-7861){\circle*{50}}
\put(9901,-7861){\circle*{50}}
\put(10201,-7861){\circle*{50}}
\put(7801,-7861){\circle*{250}}
\special{ps: gsave 0 0 0 setrgbcolor}\put(1801,-7861){\line( 1, 0){7500}}
\special{ps: grestore}\put(6901,-7711){\makebox(0,0)[lb]{\smash{\SetFigFont{8}{16.8}{\rmdefault}{\mddefault}{\updefault}\special{ps: gsave 0 0 0 setrgbcolor}+\special{ps: grestore}}}}
\put(3901,-7711){\makebox(0,0)[lb]{\smash{\SetFigFont{8}{16.8}{\rmdefault}{\mddefault}{\updefault}\special{ps: gsave 0 0 0 setrgbcolor}+\special{ps: grestore}}}}
\put(8401,-7711){\makebox(0,0)[lb]{\smash{\SetFigFont{8}{16.8}{\rmdefault}{\mddefault}{\updefault}\special{ps: gsave 0 0 0 setrgbcolor}-\special{ps: grestore}}}}
\put(5401,-7711){\makebox(0,0)[lb]{\smash{\SetFigFont{8}{16.8}{\rmdefault}{\mddefault}{\updefault}\special{ps: gsave 0 0 0 setrgbcolor}-\special{ps: grestore}}}}
\put(2401,-7711){\makebox(0,0)[lb]{\smash{\SetFigFont{8}{16.8}{\rmdefault}{\mddefault}{\updefault}\special{ps: gsave 0 0 0 setrgbcolor}-\special{ps: grestore}}}}
\end{picture}
\end{center}
  
A perturbative flow then corresponds to deleting a set of positive
links, 
\begin{center}
  \setlength{\unitlength}{1000sp}%
\begingroup\makeatletter\ifx\SetFigFont\undefined%
\gdef\SetFigFont#1#2#3#4#5{%
  \reset@font\fontsize{#1}{#2pt}%
  \fontfamily{#3}\fontseries{#4}\fontshape{#5}%
  \selectfont}%
\fi\endgroup%
\begin{picture}(18480,3560)(-3739,-11001)
\thicklines
\put(3301,-7861){\circle*{250}}
\put(1801,-7861){\circle*{250}}
\put(4801,-7861){\circle*{250}}
\put(6301,-7861){\circle*{250}}
\put(9301,-7861){\circle*{250}}
\put(9601,-7861){\circle*{50}}
\put(9901,-7861){\circle*{50}}
\put(10201,-7861){\circle*{50}}
\put(7801,-7861){\circle*{250}}
\put(2401,-9361){\circle*{250}}
\put(3901,-9361){\circle*{250}}
\put(4201,-9361){\circle*{50}}
\put(4501,-9361){\circle*{50}}
\put(4801,-9361){\circle*{50}}
\put(901,-9361){\circle*{250}}
\put(-599,-9361){\circle*{250}}
\put(-2099,-9361){\circle*{250}}
\put(-3599,-9361){\circle*{250}}
\put(1801,-10861){\circle*{250}}
\put(3301,-10861){\circle*{250}}
\put(3601,-10861){\circle*{50}}
\put(3901,-10861){\circle*{50}}
\put(4201,-10861){\circle*{50}}
\put(4501,-10861){\circle*{50}}
\put(4801,-10861){\circle*{250}}
\put(6301,-10861){\circle*{250}}
\put(6601,-10861){\circle*{50}}
\put(6901,-10861){\circle*{50}}
\put(7201,-10861){\circle*{50}}
\put(7501,-10861){\circle*{50}}
\put(7801,-10861){\circle*{250}}
\put(9301,-10861){\circle*{250}}
\put(9601,-10861){\circle*{50}}
\put(9901,-10861){\circle*{50}}
\put(10201,-10861){\circle*{50}}
\put(6301,-9361){\circle*{250}}
\put(7801,-9361){\circle*{250}}
\put(9301,-9361){\circle*{250}}
\put(10801,-9361){\circle*{250}}
\put(12301,-9361){\circle*{250}}
\put(13801,-9361){\circle*{250}}
\put(14101,-9361){\circle*{50}}
\put(14401,-9361){\circle*{50}}
\put(14701,-9361){\circle*{50}}
\put(2101,-9361){\circle*{50}}
\put(1801,-9361){\circle*{50}}
\put(1501,-9361){\circle*{50}}
\put(1201,-9361){\circle*{50}}
\put(8101,-9361){\circle*{50}}
\put(8401,-9361){\circle*{50}}
\put(8701,-9361){\circle*{50}}
\put(9001,-9361){\circle*{50}}
\special{ps: gsave 0 0 0 setrgbcolor}\put(1801,-7861){\line( 1, 0){7500}}
\special{ps: grestore}\special{ps: gsave 0 0 0 setrgbcolor}\put(3001,-8161){\vector(-1,-2){300}}
\special{ps: grestore}\special{ps: gsave 0 0 0 setrgbcolor}\put(2761,-9661){\vector( 1,-2){300}}
\special{ps: grestore}\special{ps: gsave 0 0 0 setrgbcolor}\put(1801,-10861){\line( 1, 0){1500}}
\special{ps: grestore}\special{ps: gsave 0 0 0 setrgbcolor}\put(4801,-10861){\line( 1, 0){1500}}
\special{ps: grestore}\special{ps: gsave 0 0 0 setrgbcolor}\put(7801,-10861){\line( 1, 0){1500}}
\special{ps: grestore}\special{ps: gsave 0 0 0 setrgbcolor}\put(8398,-8155){\vector( 1,-2){300}}
\special{ps: grestore}\special{ps: gsave 0 0 0 setrgbcolor}\put(8701,-9661){\vector(-1,-2){300}}
\special{ps: grestore}\special{ps: gsave 0 0 0 setrgbcolor}\put(2401,-9361){\line( 1, 0){1500}}
\special{ps: grestore}\special{ps: gsave 0 0 0 setrgbcolor}\put(-3599,-9361){\line( 1, 0){4500}}
\special{ps: grestore}\special{ps: gsave 0 0 0 setrgbcolor}\put(6451,-9361){\line( 1, 0){1500}}
\special{ps: grestore}\special{ps: gsave 0 0 0 setrgbcolor}\put(9301,-9361){\line( 1, 0){4500}}
\special{ps: grestore}\put(6901,-7711){\makebox(0,0)[lb]{\smash{\SetFigFont{8}{16.8}{\rmdefault}{\mddefault}{\updefault}\special{ps: gsave 0 0 0 setrgbcolor}+\special{ps: grestore}}}}
\put(3901,-7711){\makebox(0,0)[lb]{\smash{\SetFigFont{8}{16.8}{\rmdefault}{\mddefault}{\updefault}\special{ps: gsave 0 0 0 setrgbcolor}+\special{ps: grestore}}}}
\put(8401,-7711){\makebox(0,0)[lb]{\smash{\SetFigFont{8}{16.8}{\rmdefault}{\mddefault}{\updefault}\special{ps: gsave 0 0 0 setrgbcolor}-\special{ps: grestore}}}}
\put(5401,-7711){\makebox(0,0)[lb]{\smash{\SetFigFont{8}{16.8}{\rmdefault}{\mddefault}{\updefault}\special{ps: gsave 0 0 0 setrgbcolor}-\special{ps: grestore}}}}
\put(2401,-7711){\makebox(0,0)[lb]{\smash{\SetFigFont{8}{16.8}{\rmdefault}{\mddefault}{\updefault}\special{ps: gsave 0 0 0 setrgbcolor}-\special{ps: grestore}}}}
\end{picture}
\end{center}
We now turn to our boundary changing flows.  For this we consider
a superposition of $(1,s)$ boundaries, and represent the boundary
changing operators as vertical links joining the end nodes of the
boundaries, for example $(1,1) \oplus (1,3) \oplus (1.5)$,
\begin{center}
  \setlength{\unitlength}{1000sp}%
\begingroup\makeatletter\ifx\SetFigFont\undefined%
\gdef\SetFigFont#1#2#3#4#5{%
  \reset@font\fontsize{#1}{#2pt}%
  \fontfamily{#3}\fontseries{#4}\fontshape{#5}%
  \selectfont}%
\fi\endgroup%
\begin{picture}(8580,1480)(1661,-9201)
\thicklines
\put(3301,-7861){\circle*{250}}
\put(4801,-7861){\circle*{250}}
\put(6301,-7861){\circle*{250}}
\put(9301,-7861){\circle*{250}}
\put(9601,-7861){\circle*{50}}
\put(9901,-7861){\circle*{50}}
\put(10201,-7861){\circle*{50}}
\put(7801,-7861){\circle*{250}}
\put(9901,-9061){\circle*{50}}
\put(10201,-9061){\circle*{50}}
\put(9601,-9061){\circle*{50}}
\put(4501,-9061){\circle*{50}}
\put(4201,-9061){\circle*{50}}
\put(3901,-9061){\circle*{50}}
\put(3601,-9061){\circle*{50}}
\put(1801,-8461){\circle*{250}}
\put(3301,-8461){\circle*{250}}
\put(6301,-8461){\circle*{250}}
\put(4801,-8461){\circle*{250}}
\put(7801,-8461){\circle*{250}}
\put(9301,-8461){\circle*{250}}
\put(1801,-9061){\circle*{250}}
\put(3301,-9061){\circle*{250}}
\put(4801,-9061){\circle*{250}}
\put(6301,-9061){\circle*{250}}
\put(7801,-9061){\circle*{250}}
\put(9301,-9061){\circle*{250}}
\put(6601,-8461){\circle*{50}}
\put(6901,-8461){\circle*{50}}
\put(7201,-8461){\circle*{50}}
\put(7501,-8461){\circle*{50}}
\put(9601,-8461){\circle*{50}}
\put(9901,-8461){\circle*{50}}
\put(10201,-8461){\circle*{50}}
\put(1801,-7861){\circle*{250}}
\put(3601,-7861){\circle*{50}}
\put(3901,-7861){\circle*{50}}
\put(4201,-7861){\circle*{50}}
\put(4501,-7861){\circle*{50}}
\put(5101,-7861){\circle*{50}}
\put(5401,-7861){\circle*{50}}
\put(5776,-7861){\circle*{50}}
\put(6001,-7861){\circle*{50}}
\put(6601,-7861){\circle*{50}}
\put(6976,-7861){\circle*{50}}
\put(7276,-7861){\circle*{50}}
\put(7501,-7861){\circle*{50}}
\put(8101,-7861){\circle*{50}}
\put(8401,-7861){\circle*{50}}
\put(8701,-7861){\circle*{50}}
\put(9001,-7861){\circle*{50}}
\put(9001,-8461){\circle*{50}}
\put(8701,-8461){\circle*{50}}
\put(8401,-8461){\circle*{50}}
\put(8101,-8461){\circle*{50}}
\put(6001,-9061){\circle*{50}}
\put(5701,-9061){\circle*{50}}
\put(5401,-9061){\circle*{50}}
\put(5101,-9061){\circle*{50}}
\put(3001,-9061){\circle*{50}}
\put(2701,-9061){\circle*{50}}
\put(2401,-9061){\circle*{50}}
\put(2101,-9061){\circle*{50}}
\put(2101,-8461){\circle*{50}}
\put(2401,-8461){\circle*{50}}
\put(2701,-8461){\circle*{50}}
\put(3001,-8461){\circle*{50}}
\special{ps: gsave 0 0 0 setrgbcolor}\put(3301,-8461){\line( 1, 0){1500}}
\special{ps: grestore}\special{ps: gsave 0 0 0 setrgbcolor}\put(4801,-8461){\line( 1, 0){1500}}
\special{ps: grestore}\special{ps: gsave 0 0 0 setrgbcolor}\put(6301,-9061){\line( 1, 0){3000}}
\special{ps: grestore}\special{ps: gsave 0 0 0 setrgbcolor}\put(1801,-7861){\line( 1, 0){1500}}
\special{ps: grestore}\special{ps: gsave 0 0 0 setrgbcolor}\put(3376,-7861){\line( 0,-1){600}}
\special{ps: grestore}\special{ps: gsave 0 0 0 setrgbcolor}\put(3226,-7861){\line( 0,-1){600}}
\special{ps: grestore}\special{ps: gsave 0 0 0 setrgbcolor}\put(6226,-8461){\line( 0,-1){600}}
\special{ps: grestore}\special{ps: gsave 0 0 0 setrgbcolor}\put(6376,-8461){\line( 0,-1){600}}
\special{ps: grestore}\end{picture}
\end{center}
Boundary changing flows then correspond to deleting these links and
fusing the subgraphs together,
\begin{center}
  \setlength{\unitlength}{1000sp}%
\begingroup\makeatletter\ifx\SetFigFont\undefined%
\gdef\SetFigFont#1#2#3#4#5{%
  \reset@font\fontsize{#1}{#2pt}%
  \fontfamily{#3}\fontseries{#4}\fontshape{#5}%
  \selectfont}%
\fi\endgroup%
\begin{picture}(18180,5093)(-3439,-12814)
\thicklines
\put(3301,-7861){\circle*{250}}
\put(4801,-7861){\circle*{250}}
\put(6301,-7861){\circle*{250}}
\put(9301,-7861){\circle*{250}}
\put(9601,-7861){\circle*{50}}
\put(9901,-7861){\circle*{50}}
\put(10201,-7861){\circle*{50}}
\put(7801,-7861){\circle*{250}}
\put(9901,-9061){\circle*{50}}
\put(10201,-9061){\circle*{50}}
\put(9601,-9061){\circle*{50}}
\put(4501,-9061){\circle*{50}}
\put(4201,-9061){\circle*{50}}
\put(3901,-9061){\circle*{50}}
\put(3601,-9061){\circle*{50}}
\put(1801,-8461){\circle*{250}}
\put(3301,-8461){\circle*{250}}
\put(6301,-8461){\circle*{250}}
\put(4801,-8461){\circle*{250}}
\put(7801,-8461){\circle*{250}}
\put(9301,-8461){\circle*{250}}
\put(1801,-9061){\circle*{250}}
\put(3301,-9061){\circle*{250}}
\put(4801,-9061){\circle*{250}}
\put(6301,-9061){\circle*{250}}
\put(7801,-9061){\circle*{250}}
\put(9301,-9061){\circle*{250}}
\put(6601,-8461){\circle*{50}}
\put(6901,-8461){\circle*{50}}
\put(7201,-8461){\circle*{50}}
\put(7501,-8461){\circle*{50}}
\put(9601,-8461){\circle*{50}}
\put(9901,-8461){\circle*{50}}
\put(10201,-8461){\circle*{50}}
\put(1801,-7861){\circle*{250}}
\put(3601,-7861){\circle*{50}}
\put(3901,-7861){\circle*{50}}
\put(4201,-7861){\circle*{50}}
\put(4501,-7861){\circle*{50}}
\put(5101,-7861){\circle*{50}}
\put(5401,-7861){\circle*{50}}
\put(5776,-7861){\circle*{50}}
\put(6001,-7861){\circle*{50}}
\put(6601,-7861){\circle*{50}}
\put(6976,-7861){\circle*{50}}
\put(7276,-7861){\circle*{50}}
\put(7501,-7861){\circle*{50}}
\put(8101,-7861){\circle*{50}}
\put(8401,-7861){\circle*{50}}
\put(8701,-7861){\circle*{50}}
\put(9001,-7861){\circle*{50}}
\put(9001,-8461){\circle*{50}}
\put(8701,-8461){\circle*{50}}
\put(8401,-8461){\circle*{50}}
\put(8101,-8461){\circle*{50}}
\put(6001,-9061){\circle*{50}}
\put(5701,-9061){\circle*{50}}
\put(5401,-9061){\circle*{50}}
\put(5101,-9061){\circle*{50}}
\put(3001,-9061){\circle*{50}}
\put(2701,-9061){\circle*{50}}
\put(2401,-9061){\circle*{50}}
\put(2101,-9061){\circle*{50}}
\put(2101,-8461){\circle*{50}}
\put(2401,-8461){\circle*{50}}
\put(2701,-8461){\circle*{50}}
\put(3001,-8461){\circle*{50}}
\put(4201,-10561){\circle*{50}}
\put(4476,-10561){\circle*{50}}
\put(4501,-11161){\circle*{50}}
\put(4176,-11161){\circle*{50}}
\put(4776,-10561){\circle*{50}}
\put(4801,-11161){\circle*{50}}
\put(3901,-11161){\circle*{250}}
\put(3901,-10561){\circle*{250}}
\put(3601,-10561){\circle*{50}}
\put(3301,-10561){\circle*{50}}
\put(3001,-10561){\circle*{50}}
\put(2701,-10561){\circle*{50}}
\put(2101,-10561){\circle*{50}}
\put(1801,-10561){\circle*{50}}
\put(1501,-10561){\circle*{50}}
\put(1201,-10561){\circle*{50}}
\put(601,-11161){\circle*{50}}
\put(301,-11161){\circle*{50}}
\put(2401,-10561){\circle*{250}}
\put(2401,-11161){\circle*{250}}
\put(901,-10561){\circle*{250}}
\put(901,-11161){\circle*{250}}
\put(-599,-11161){\circle*{250}}
\put(-599,-10561){\circle*{250}}
\put(-2099,-10561){\circle*{250}}
\put(-2099,-11161){\circle*{250}}
\put(-3299,-11161){\circle*{250}}
\put(-3299,-10561){\circle*{250}}
\put(  1,-11161){\circle*{50}}
\put(-224,-11161){\circle*{50}}
\put(-824,-11161){\circle*{50}}
\put(-1199,-11161){\circle*{50}}
\put(-1499,-11161){\circle*{50}}
\put(-1799,-11161){\circle*{50}}
\put(-2399,-11161){\circle*{50}}
\put(-2699,-11161){\circle*{50}}
\put(-2999,-11161){\circle*{50}}
\put(6301,-10561){\circle*{250}}
\put(6301,-11161){\circle*{250}}
\put(7788,-10574){\circle*{250}}
\put(7801,-11161){\circle*{250}}
\put(6601,-11161){\circle*{50}}
\put(6901,-11161){\circle*{50}}
\put(7201,-11161){\circle*{50}}
\put(7501,-11161){\circle*{50}}
\put(8101,-10561){\circle*{50}}
\put(8401,-10561){\circle*{50}}
\put(8701,-10561){\circle*{50}}
\put(9001,-10561){\circle*{50}}
\put(9601,-10561){\circle*{50}}
\put(9901,-10561){\circle*{50}}
\put(10201,-10561){\circle*{50}}
\put(10501,-10561){\circle*{50}}
\put(11101,-10561){\circle*{50}}
\put(11401,-10561){\circle*{50}}
\put(11701,-10561){\circle*{50}}
\put(12001,-10561){\circle*{50}}
\put(12601,-10561){\circle*{50}}
\put(12901,-10561){\circle*{50}}
\put(13201,-10561){\circle*{50}}
\put(9301,-10561){\circle*{250}}
\put(9301,-11161){\circle*{250}}
\put(10801,-11161){\circle*{250}}
\put(10801,-10561){\circle*{250}}
\put(12301,-10561){\circle*{250}}
\put(1801,-12661){\circle*{250}}
\put(3288,-12674){\circle*{250}}
\put(7858,-12635){\circle*{250}}
\put(9283,-12635){\circle*{250}}
\put(9601,-12661){\circle*{50}}
\put(9901,-12661){\circle*{50}}
\put(10201,-12661){\circle*{50}}
\put(6301,-12661){\circle*{250}}
\put(4801,-12661){\circle*{250}}
\put(13501,-10561){\circle*{50}}
\put(14101,-10561){\circle*{50}}
\put(14401,-10561){\circle*{50}}
\put(14701,-10561){\circle*{50}}
\put(14101,-11161){\circle*{50}}
\put(14401,-11161){\circle*{50}}
\put(14701,-11161){\circle*{50}}
\put(13801,-11161){\circle*{250}}
\put(12301,-11161){\circle*{250}}
\put(13801,-10561){\circle*{250}}
\special{ps: gsave 0 0 0 setrgbcolor}\put(3301,-8461){\line( 1, 0){1500}}
\special{ps: grestore}\special{ps: gsave 0 0 0 setrgbcolor}\put(4801,-8461){\line( 1, 0){1500}}
\special{ps: grestore}\special{ps: gsave 0 0 0 setrgbcolor}\put(6301,-9061){\line( 1, 0){3000}}
\special{ps: grestore}\special{ps: gsave 0 0 0 setrgbcolor}\put(1801,-7861){\line( 1, 0){1500}}
\special{ps: grestore}\special{ps: gsave 0 0 0 setrgbcolor}\put(3376,-7861){\line( 0,-1){600}}
\special{ps: grestore}\special{ps: gsave 0 0 0 setrgbcolor}\put(3226,-7861){\line( 0,-1){600}}
\special{ps: grestore}\special{ps: gsave 0 0 0 setrgbcolor}\put(6226,-8461){\line( 0,-1){600}}
\special{ps: grestore}\special{ps: gsave 0 0 0 setrgbcolor}\put(6376,-8461){\line( 0,-1){600}}
\special{ps: grestore}\special{ps: gsave 0 0 0 setrgbcolor}\put(8401,-9361){\vector( 1,-3){202.500}}
\special{ps: grestore}\special{ps: gsave 0 0 0 setrgbcolor}\put(901,-11161){\line( 1, 0){3075}}
\special{ps: grestore}\special{ps: gsave 0 0 0 setrgbcolor}\put(-3299,-10561){\line( 1, 0){4200}}
\special{ps: grestore}\special{ps: gsave 0 0 0 setrgbcolor}\put(826,-10561){\line( 0,-1){600}}
\special{ps: grestore}\special{ps: gsave 0 0 0 setrgbcolor}\put(976,-10561){\line( 0,-1){600}}
\special{ps: grestore}\special{ps: gsave 0 0 0 setrgbcolor}\put(6301,-10561){\line( 1, 0){1500}}
\special{ps: grestore}\special{ps: gsave 0 0 0 setrgbcolor}\put(7726,-10561){\line( 0,-1){600}}
\special{ps: grestore}\special{ps: gsave 0 0 0 setrgbcolor}\put(7876,-10561){\line( 0,-1){600}}
\special{ps: grestore}\special{ps: gsave 0 0 0 setrgbcolor}\put(7801,-11161){\line( 1, 0){6000}}
\special{ps: grestore}\special{ps: gsave 0 0 0 setrgbcolor}\put(2993,-9356){\vector(-1,-2){270}}
\special{ps: grestore}\special{ps: gsave 0 0 0 setrgbcolor}\put(2843,-11540){\vector( 1,-2){270}}
\special{ps: grestore}\special{ps: gsave 0 0 0 setrgbcolor}\put(8634,-11539){\vector(-1,-3){202.500}}
\special{ps: grestore}\special{ps: gsave 0 0 0 setrgbcolor}\put(1801,-12661){\line( 1, 0){7500}}
\special{ps: grestore}\end{picture}
\end{center} 

One can see that in general, all the possible perturbative end points of flows
from a superposition of $(1,s)$ are represented in the subgraphs
generated by successive deletions and fusings.  Furthermore, it is our
belief that the space of possible perturbative endpoints is
represented by this procedure:  A perturbative flow from $a \to b$
exists if and only if the graph of $b$ is a subgraph of $a$ obtained
by fusing and/or by deleting positive links.  \\

As an example, we consider the case of $(1,p-1)\oplus (1,p+1)$.  Using
the above rules we construct the following diagram.
\begin{center}
  \setlength{\unitlength}{1000sp}%
\begingroup\makeatletter\ifx\SetFigFont\undefined%
\gdef\SetFigFont#1#2#3#4#5{%
  \reset@font\fontsize{#1}{#2pt}%
  \fontfamily{#3}\fontseries{#4}\fontshape{#5}%
  \selectfont}%
\fi\endgroup%
\begin{picture}(17496,6764)(-2447,-8301)
\thicklines
\put(3001,-2161){\circle*{50}}
\put(3301,-2161){\circle*{50}}
\put(3601,-2161){\circle*{50}}
\put(3914,-2174){\circle*{250}}
\put(5401,-2161){\circle*{250}}
\put(6901,-2161){\circle*{250}}
\put(6901,-2761){\circle*{250}}
\put(5401,-2761){\circle*{250}}
\put(3901,-2761){\circle*{250}}
\put(8401,-2161){\circle*{250}}
\put(8401,-2761){\circle*{250}}
\put(9901,-2761){\circle*{250}}
\put(9901,-2161){\circle*{250}}
\put(3001,-2761){\circle*{50}}
\put(3301,-2761){\circle*{50}}
\put(3601,-2761){\circle*{50}}
\put(4201,-2761){\circle*{50}}
\put(4501,-2761){\circle*{50}}
\put(4801,-2761){\circle*{50}}
\put(5101,-2761){\circle*{50}}
\put(5701,-2761){\circle*{50}}
\put(6001,-2761){\circle*{50}}
\put(6301,-2761){\circle*{50}}
\put(6601,-2761){\circle*{50}}
\put(7201,-2161){\circle*{50}}
\put(7501,-2161){\circle*{50}}
\put(7801,-2161){\circle*{50}}
\put(8101,-2161){\circle*{50}}
\put(8701,-2161){\circle*{50}}
\put(9001,-2161){\circle*{50}}
\put(9301,-2161){\circle*{50}}
\put(9601,-2161){\circle*{50}}
\put(10201,-2161){\circle*{50}}
\put(10501,-2161){\circle*{50}}
\put(10801,-2161){\circle*{50}}
\put(10201,-2761){\circle*{50}}
\put(10501,-2761){\circle*{50}}
\put(10801,-2761){\circle*{50}}
\put(7801,-4861){\circle*{50}}
\put(7501,-4861){\circle*{50}}
\put(7501,-4261){\circle*{50}}
\put(7801,-4261){\circle*{50}}
\put(7201,-4261){\circle*{50}}
\put(7201,-4861){\circle*{50}}
\put(8101,-4261){\circle*{250}}
\put(8101,-4861){\circle*{250}}
\put(9601,-4261){\circle*{250}}
\put(9601,-4861){\circle*{250}}
\put(11101,-4861){\circle*{250}}
\put(11101,-4261){\circle*{250}}
\put(12601,-4261){\circle*{250}}
\put(12601,-4861){\circle*{250}}
\put(14101,-4861){\circle*{250}}
\put(14101,-4261){\circle*{250}}
\put(8401,-4861){\circle*{50}}
\put(8701,-4861){\circle*{50}}
\put(9001,-4861){\circle*{50}}
\put(9301,-4861){\circle*{50}}
\put(9901,-4861){\circle*{50}}
\put(10276,-4861){\circle*{50}}
\put(10501,-4861){\circle*{50}}
\put(10801,-4861){\circle*{50}}
\put(11701,-4261){\circle*{50}}
\put(12001,-4261){\circle*{50}}
\put(12901,-4261){\circle*{50}}
\put(13201,-4261){\circle*{50}}
\put(13501,-4261){\circle*{50}}
\put(13801,-4261){\circle*{50}}
\put(11401,-4261){\circle*{50}}
\put(12301,-4261){\circle*{50}}
\put(14401,-4261){\circle*{50}}
\put(14701,-4261){\circle*{50}}
\put(15001,-4261){\circle*{50}}
\put(15001,-4861){\circle*{50}}
\put(14701,-4861){\circle*{50}}
\put(14401,-4861){\circle*{50}}
\put(-1499,-4561){\circle*{50}}
\put(-1199,-4561){\circle*{50}}
\put(-899,-4561){\circle*{50}}
\put(-599,-4561){\circle*{250}}
\put(901,-4561){\circle*{250}}
\put(2401,-4561){\circle*{250}}
\put(5401,-4561){\circle*{250}}
\put(3901,-4561){\circle*{250}}
\put(5701,-4561){\circle*{50}}
\put(6001,-4561){\circle*{50}}
\put(6301,-4561){\circle*{50}}
\put(3001,-6361){\circle*{250}}
\put(4501,-6361){\circle*{250}}
\put(5401,-6361){\circle*{50}}
\put(5101,-6361){\circle*{50}}
\put(4801,-6361){\circle*{50}}
\put(1501,-6361){\circle*{250}}
\put(  1,-6361){\circle*{250}}
\put(-1499,-6361){\circle*{250}}
\put(-2399,-6361){\circle*{50}}
\put(-2099,-6361){\circle*{50}}
\put(-1799,-6361){\circle*{50}}
\put(8976,-4261){\circle*{50}}
\put(9301,-4261){\circle*{50}}
\put(8376,-4261){\circle*{50}}
\put(8701,-4261){\circle*{50}}
\put(7201,-6361){\circle*{250}}
\put(7501,-6361){\circle*{50}}
\put(7801,-6361){\circle*{50}}
\put(8101,-6361){\circle*{50}}
\put(8401,-6361){\circle*{50}}
\put(6876,-6361){\circle*{50}}
\put(6576,-6361){\circle*{50}}
\put(6276,-6361){\circle*{50}}
\put(8701,-6361){\circle*{250}}
\put(10201,-6361){\circle*{250}}
\put(11701,-6361){\circle*{250}}
\put(13201,-6361){\circle*{250}}
\put(13476,-6361){\circle*{50}}
\put(13801,-6361){\circle*{50}}
\put(14076,-6361){\circle*{50}}
\put(9901,-8161){\circle*{250}}
\put(8401,-8161){\circle*{250}}
\put(6901,-8161){\circle*{250}}
\put(5401,-8161){\circle*{250}}
\put(4501,-8161){\circle*{50}}
\put(4801,-8161){\circle*{50}}
\put(5101,-8161){\circle*{50}}
\put(7201,-8161){\circle*{50}}
\put(7501,-8161){\circle*{50}}
\put(7801,-8161){\circle*{50}}
\put(8101,-8161){\circle*{50}}
\put(10201,-8161){\circle*{50}}
\put(10501,-8161){\circle*{50}}
\put(10801,-8161){\circle*{50}}
\put(4201,-8161){\circle*{50}}
\put(3901,-8161){\circle*{250}}
\put(3601,-8161){\circle*{50}}
\put(3301,-8161){\circle*{50}}
\put(2976,-8161){\circle*{50}}
\put(2076,-6361){\circle*{50}}
\put(1776,-6361){\circle*{50}}
\put(2676,-6361){\circle*{50}}
\put(2401,-6361){\circle*{50}}
\special{ps: gsave 0 0 0 setrgbcolor}\put(3901,-2161){\line( 1, 0){3000}}
\special{ps: grestore}\special{ps: gsave 0 0 0 setrgbcolor}\put(6901,-2761){\line( 1, 0){3000}}
\special{ps: grestore}\special{ps: gsave 0 0 0 setrgbcolor}\put(6976,-2161){\line( 0,-1){600}}
\special{ps: grestore}\special{ps: gsave 0 0 0 setrgbcolor}\put(6826,-2161){\line( 0,-1){600}}
\special{ps: grestore}\special{ps: gsave 0 0 0 setrgbcolor}\put(4732,-3134){\vector(-2,-3){545.077}}
\special{ps: grestore}\special{ps: gsave 0 0 0 setrgbcolor}\put(9001,-6661){\vector(-1,-3){299}}
\special{ps: grestore}\special{ps: gsave 0 0 0 setrgbcolor}\put(2701,-4861){\vector(-1,-3){300}}
\special{ps: grestore}\special{ps: gsave 0 0 0 setrgbcolor}\put(6001,-4861){\vector( 2,-3){600}}
\special{ps: grestore}\special{ps: gsave 0 0 0 setrgbcolor}\put(2694,-6732){\vector( 1,-3){279.400}}
\special{ps: grestore}\special{ps: gsave 0 0 0 setrgbcolor}\put(9026,-3118){\vector( 1,-2){301.800}}
\special{ps: grestore}\special{ps: gsave 0 0 0 setrgbcolor}\put(9590,-5190){\vector(-1,-2){301.800}}
\special{ps: grestore}\special{ps: gsave 0 0 0 setrgbcolor}\put(11176,-4261){\line( 0,-1){600}}
\special{ps: grestore}\special{ps: gsave 0 0 0 setrgbcolor}\put(11026,-4261){\line( 0,-1){600}}
\special{ps: grestore}\special{ps: gsave 0 0 0 setrgbcolor}\put(-599,-4561){\line( 1, 0){6000}}
\special{ps: grestore}\special{ps: gsave 0 0 0 setrgbcolor}\put(9601,-4261){\line( 1, 0){1500}}
\special{ps: grestore}\special{ps: gsave 0 0 0 setrgbcolor}\put(11101,-4861){\line( 1, 0){3000}}
\special{ps: grestore}\special{ps: gsave 0 0 0 setrgbcolor}\put(8701,-6361){\line( 1, 0){4500}}
\special{ps: grestore}\special{ps: gsave 0 0 0 setrgbcolor}\put(5401,-8161){\line( 1, 0){1500}}
\special{ps: grestore}\special{ps: gsave 0 0 0 setrgbcolor}\put(8401,-8161){\line( 1, 0){1500}}
\special{ps: grestore}\special{ps: gsave 0 0 0 setrgbcolor}\put(3001,-6361){\line( 1, 0){1500}}
\special{ps: grestore}\special{ps: gsave 0 0 0 setrgbcolor}\put(-1499,-6361){\line( 1, 0){3000}}
\special{ps: grestore}\put(8026,-1861){\makebox(0,0)[lb]{\smash{\SetFigFont{8}{14.4}{\familydefault}{\mddefault}{\updefault}\special{ps: gsave 0 0 0 setrgbcolor}$p+1$\special{ps: grestore}}}}
\put(4951,-1861){\makebox(0,0)[lb]{\smash{\SetFigFont{8}{14.4}{\familydefault}{\mddefault}{\updefault}\special{ps: gsave 0 0 0 setrgbcolor}$p-1$\special{ps: grestore}}}}
\put(6676,-1861){\makebox(0,0)[lb]{\smash{\SetFigFont{8}{14.4}{\familydefault}{\mddefault}{\updefault}\special{ps: gsave 0 0 0 setrgbcolor}$p$\special{ps: grestore}}}}
\put(2251,-4261){\makebox(0,0)[lb]{\smash{\SetFigFont{8}{14.4}{\familydefault}{\mddefault}{\updefault}\special{ps: gsave 0 0 0 setrgbcolor}$p$\special{ps: grestore}}}}
\put(1351,-6061){\makebox(0,0)[lb]{\smash{\SetFigFont{8}{14.4}{\familydefault}{\mddefault}{\updefault}\special{ps: gsave 0 0 0 setrgbcolor}$p$\special{ps: grestore}}}}
\put(6751,-7861){\makebox(0,0)[lb]{\smash{\SetFigFont{8}{14.4}{\familydefault}{\mddefault}{\updefault}\special{ps: gsave 0 0 0 setrgbcolor}$p$\special{ps: grestore}}}}
\put(10951,-3961){\makebox(0,0)[lb]{\smash{\SetFigFont{8}{14.4}{\familydefault}{\mddefault}{\updefault}\special{ps: gsave 0 0 0 setrgbcolor}$p$\special{ps: grestore}}}}
\put(10051,-6061){\makebox(0,0)[lb]{\smash{\SetFigFont{8}{14.4}{\familydefault}{\mddefault}{\updefault}\special{ps: gsave 0 0 0 setrgbcolor}$p$\special{ps: grestore}}}}
\end{picture} \\[+20pt]
  \setlength{\unitlength}{1000sp}%
\begingroup\makeatletter\ifx\SetFigFont\undefined%
\gdef\SetFigFont#1#2#3#4#5{%
  \reset@font\fontsize{#1}{#2pt}%
  \fontfamily{#3}\fontseries{#4}\fontshape{#5}%
  \selectfont}%
\fi\endgroup%
\begin{picture}(9622,5878)(1,-8366)
\thicklines
\special{ps: gsave 0 0 0 setrgbcolor}\put(4732,-3134){\vector(-2,-3){545.077}}
\special{ps: grestore}\special{ps: gsave 0 0 0 setrgbcolor}\put(3301,-4861){\vector(-1,-1){900}}
\special{ps: grestore}\special{ps: gsave 0 0 0 setrgbcolor}\put(5101,-4861){\vector( 3,-2){1453.846}}
\special{ps: grestore}\special{ps: gsave 0 0 0 setrgbcolor}\put(2694,-6732){\vector( 1,-3){279.400}}
\special{ps: grestore}\special{ps: gsave 0 0 0 setrgbcolor}\put(9026,-3118){\vector( 1,-2){301.800}}
\special{ps: grestore}\special{ps: gsave 0 0 0 setrgbcolor}\put(8790,-4861){\vector(-1,-2){301.800}}
\special{ps: grestore}\special{ps: gsave 0 0 0 setrgbcolor}\put(7794,-6658){\vector(-1,-3){299}}
\special{ps: grestore}\put(4801,-2836){\makebox(0,0)[lb]{\smash{\SetFigFont{8}{14.4}{\familydefault}{\mddefault}{\updefault}\special{ps: gsave 0 0 0 setrgbcolor}$(1,p-1)\oplus (1,p+1)$\special{ps: grestore}}}}
\put(7426,-4486){\makebox(0,0)[lb]{\smash{\SetFigFont{8}{14.4}{\familydefault}{\mddefault}{\updefault}\special{ps: gsave 0 0 0 setrgbcolor}$(p-1,1)\oplus (1,p+1)$\special{ps: grestore}}}}
\put(7276,-6286){\makebox(0,0)[lb]{\smash{\SetFigFont{8}{14.4}{\familydefault}{\mddefault}{\updefault}\special{ps: gsave 0 0 0 setrgbcolor}$(p,2)$\special{ps: grestore}}}}
\put(  1,-6361){\makebox(0,0)[lb]{\smash{\SetFigFont{8}{14.4}{\familydefault}{\mddefault}{\updefault}\special{ps: gsave 0 0 0 setrgbcolor}$(1,p-1)\oplus (p+1,1)$\special{ps: grestore}}}}
\put(3376,-4561){\makebox(0,0)[lb]{\smash{\SetFigFont{8}{14.4}{\familydefault}{\mddefault}{\updefault}\special{ps: gsave 0 0 0 setrgbcolor}$(2,p)$\special{ps: grestore}}}}
\put(3001,-8236){\makebox(0,0)[lb]{\smash{\SetFigFont{8}{14.4}{\familydefault}{\mddefault}{\updefault}\special{ps: gsave 0 0 0 setrgbcolor}$(p-1,1)\oplus (p+1,1)$\special{ps: grestore}}}}
\end{picture}
\end{center}
Thus we predict that the $(2,p)$ boundary contains flows
to $(p,2)$ and $(1,p-1)\oplus (p+1,1)$ while $(p-1,1)\oplus (1,p+1)$
contains a $\phi_{24}$ boundary changing flow to $(p,2)$ in agreement with the numerical studies of section \ref{sec:nfrombex2}.\\

Maybe we can go further.  TCSA results \cite{Wat00} suggest that
in the non-perturbative direction the $\phi_{13}$ perturbation of a
Cardy boundary condition flowed to,
\begin{align}
  (r,s) \to \oplus_{i=1}^{\min{\{r,s-1 \}}} (r+s-2i,1) \; ,
\end{align}
see also \cite{Rec00}.  In our diagrammatic rules, this corresponds to deleting all the
negative links.  Furthermore in \cite{Gra00,Gra01}, it was observed that perturbations by
$\phi_{rr}$ fields can act as ``projectors'', flowing onto one of the
boundaries $(1,s+2t-2)$ or $(s+2t-2,1)$ for some $t=1,\ldots,n$.
In  our diagrammatic rules, this can again be represented by
projecting onto a subgraph.

All in all we feel the evidence is strong that the space of end points
of general perturbations of a $(r,s)$ boundary condition are described
succinctly by the subgraphs of its associated graph.

\resection{Conclusions}

In conclusion, we have studied perturbations of superpositions of
Cardy boundary conditions within perturbation theory.  We showed how
to study perturbations by $\phi_{r,r+2}$ fields on general
superpositions of boundary conditions.  For two examples,
boundary changing $\phi_{13}$ operators and perturbations by
$\phi_{35}$, we explicitly constructed the space of perturbative flows
and studied the fixed points.  We noted that by a suitable choice
of superposition of $(1,s)$ boundary conditions, one could construct
an arbitrary Cardy boundary condition and study its space of flows
from the field theory on the superposition.   Furthermore, we showed
that to lowest order in the examples considered, the $\beta$-functions
describing both the superposition and the Cardy boundary were identical
after a redefinition of fields.  We then went on and conjectured a
diagrammatic representation for the space of renormalisation group fixed points of a Cardy boundary condition. \\

A lot of work still needs to be done.
\begin{itemize}
\item
  The diagrammatic rules need to be proven for the perturbative flows as well as more checks carried out in the non-perturbative case.
\item
  Can one represent more general boundary changing flows diagrammatically?  In particular is there a lattice realisation for a superposition of boundary conditions?  Does there exist a lattice realisation of these flows?
\item
  Are any of these flows integrable?
\item
  What of $\phi_{r,r \pm1}$?  These fields have a character all of their own that has yet to be investigated. 
\item
  Lattice realisations of the D and E Series minimal models are also
available, are their boundary flows also represented in such a manor?
\end{itemize}  

Note added.  After the completion of this work, we became aware of \cite{Fre01} where some of the results of this paper were found by other means.

\subsection*{Acknowledgments}

The author thanks Ingo Runkel and James Drummond for helpful discussions and support.  He also thanks S.~Fredenhagen, A.~Recknagel and V.~Schomerus for comments and correspondance as well as IPAM-UCLA for their hospitality during the latter stages of this work.
 Finally, he would like to thank Gerard Watts whose help has been indispensable, thanks.  Supported by an EPSRC studentship. 

\appendix
\section*{Appendices }
\refstepcounter{section}
\label{app}

\subsection{Structure Constants for $\phi_{13}$}
\label{sec:sc13}

Due to the constraint $\bc abaii\One \langle \One_a \rangle = \bc
babii\One \langle \One_b \rangle$, it is not possible to set both $\bc
abaii\One$ and $\bc babii\One$ equal to one.  Instead we define an ordering
on pairs $a_i=(r_i,s_i)$ whereby $a_1<a_2$ if $r_1<r_2$, and if $r_1=r_2$
then $a_1<a_2$ if $s_1<s_2$.  We define,
\begin{align}
  \bc abaii\One = 
  \begin{cases} 
    1  \qquad \qquad \qquad \text{ if  } a \le b \; , \\
    \langle \One_b \rangle / \langle \One_a
    \rangle  \qquad \text{otherwise.}
  \end{cases} 
\end{align}
The leading behaviours of the structure constants for $\phi_{13}$ fields in the $m \to \infty$ limit are as follows, \\

$a=(r,1)$, $b=(r,3)$
\begin{align}
  \bc aaa... = \bc aba... = \bc aab... =  \bc aaa..\One = 0 \; , \notag
\end{align}
\vspace{-5ex}
\begin{alignat}{3}
  \bc bbb... &= 1 \; , & \qquad
  \bc bab... &= \frac 23 \; , & \qquad
  \bc abb... &= 2 \; , \notag \\
  \bc bbb..\One &= 1 \; , & \qquad
  \bc aba..\One &= 1 \; , & \qquad
  \bc bab..\One &= \frac 13 \; .
\end{alignat}

$a=(r,s)$, $b=(r,s+2)$, $s \ne 1$
\begin{alignat}{2}
  \bc aaa... &= \sqrt{\frac{8}{s^2-1}} \; , &
  \bc aab... &= \bc baa... = \bc aba... = -\sqrt{2\frac{s-1}{s+1}}
  \; , \notag \\
  \bc bab... &= \frac{s}{s+2}\sqrt{2\frac{s+3}{s+1}} \; , & \qquad
  \bc bba... &= \bc abb... = \sqrt{2\frac{s+3}{s+1}} \; , \notag
\end{alignat}
\begin{alignat}{3}
  \bc aaa..\One &= 1 \; , & \qquad
  \bc aba..\One = 1 \; , & \qquad
  \bc bab..\One &= \frac{s}{s+2} \; , 
\end{alignat}
where we have used a dot to denote  $\phi_{13}$.

\subsection{Structure Constants for the $(2,p)$ Boundary}
\label{sec:sc2p}

All fields on the $(2,p)$ boundary are normalised so that $\bc
...ii\One =1$.  In the case of $L_{-1} \phi_{33}$ we have,
\begin{align}
  L_{-1} \phi_{33}(x) \; L_{-1} \phi_{33}(0) 
  = 2h_{33}(1+2h_{33}) \; x^{-2-2h_{33}} \One + \ldots \; . 
\end{align}
To leading order $2h_{33}(1+2h_{33}) = \left( \tfrac{2}{m+1} \right)^2 $ and so we normalise, 
\begin{align}
  d_3 = \tfrac{m+1}{2} L_{-1} \phi_{33} \; .
\end{align}
The structure constants for the $(2,p)$ boundary to leading order as
$m \to \infty $ are,\newline
with $
  \phi = \phi_{13}, \;\; \tphi = \phi_{31}, \;\;  c_3 = \phi_{33} , \;\;
  d_3 = \tfrac{m+1}{2}  L_{-1}\phi_{33}, \;\; \psi = \phi_{35}$,
{\allowdisplaybreaks
\begin{align}
  &\bc ...{\phi}{\phi}{\phi} = \sqrt{\frac{8}{p^2-1}}\;,\;\;&
  &\bc ...{\tphi}{\tphi}{\tphi} = \sqrt{\frac{8}{3}}\;,\;\;&
  &\bc ...{\phi}{\tphi}{c} = -\sqrt{\frac{1}{3}}\;,\notag\\
  &\bc ...{c}{c}{c} = \sqrt{\frac{4}{p^2-1}}\;,\;\;&
  &\bc ...{\phi}{\psi}{\psi} = 3 \sqrt{\frac{2}{p^2-1}}\;,\;\;&
  &\bc ...{\tphi}{\psi}{\psi} = -\sqrt{\frac{2}{3}}\;,\notag\\
  &\bc ...{\phi}{c}{\psi} = - \sqrt{\frac 23\,\frac{p^2-4}{p^2-1}}\;,\;\;&
  &\bc ...{c}{\psi}{\psi} = \sqrt{\frac{9}{p^2-1}}\;,\;\;&
  &\bc ...{\psi}{\psi}{\psi} = (p^2{-}16)\sqrt{\frac 43\,\frac{1}{(p^2-4)(p^2-1)}}\;,\notag\\
  &\bc ...{\phi}{c}{c} = -\sqrt{\frac{8}{p^2-1}}\cdot\frac{1}{m+1}\;,\;\;&
  &\bc ...{\tphi}{c}{c} = \sqrt{\frac{8}{3}}\cdot\frac{1}{m+1}\;,\;\;&
  &\bc ...{c}{c}{\psi} = -\sqrt{\frac 43 \,
  \frac{p^2-4}{p^2-1}}\cdot\frac{1}{m+1}\;, \label{eq:sc2p1}
\end{align}}
where dots denote the $(2,p)$ boundary.  The structure constants
\eqref{eq:sc2p1} are
totally symmetric in all indices. \\

The structure
constants involving an odd number of descendent fields $d_3$ are all
antisymmetric in the lower two indices and so do not contribute to the
$\beta$-functions.  The remaining are symmetric and take the form,
\begin{align}
 &\bc ...{\phi}{d}{d} = -\bc ...{d}{d}{\phi} = \sqrt{\frac{2}{p^2-1}}
 \; , \notag \\
 &\bc ...{\tphi}{d}{d} = -\bc ...{d}{d}{\tphi} = \sqrt{\frac{2}{3}}
 \; , \notag \\
 &\bc ...{\psi}{d}{d} = -\bc ...{d}{d}{\psi} = 
 \sqrt{\frac{4}{3} \frac{p^2-4}{p^2-1}} \; .
\end{align}

\subsection{Solving the $\beta$-functions}
\label{sec:solbeta}
  
We will attempt to find a solution to the $\beta$-function equations,
\begin{align}
  L\frac{d\lam _i}{dL} = y_i \lam _i + \sum_{j,k} C_{jk}{}^i \lam _j \lam _k \; , \qquad \lam_i^b = \lam_i(\eps) \; ,
\end{align}
of the form
\begin{align}
  \lam ^b_i = \sum_j a_j^i(L)\lam _j(L) + \sum_{j,k} b^i_{jk}(L) \lam _j(L) \lam _k(L) \; .
\end{align}
\newline

First differentiate with respect to $L$ and
substitute in expression for the $\beta$-function.
\begin{align}
  0 = \sum_j \{ L\frac{da_j^i}{dL} \lam _j +a_j^i ( y_j\lam _j +
  \sum_{k,l} C_{kl}{}^j \lam _k \lam _l ) \}
  +\sum_{j,k} \{ L\frac{db^i_{jk}}{dL} \lam _j \lam _k + (y_j + y_k)
  b^i_{jk} \lam _j \lam _k \} \; .
\end{align}
Equating coefficients one obtains the following ODE's,
\begin{align}
  0 &= L\frac{da_j^i}{dL} + y_j a_j^i \; , \\
  0 &= L\frac{db^i_{jk}}{dL} + (y_j + y_k) b^i_{jk} + \sum _l C_{jk}{}^l
  a_l^i \; ,
\end{align}
which using the boundary condition become,
\begin{align}
  a_j^i &= \delta_j^i \left( \tfrac{\eps}{L} \right)^{y_j} \; , \\
  b^i_{jk} &=
  \begin{cases}
     \frac{1}{y_i - y_j - y_k}C_{jk}{}^i [  \left( \tfrac{\eps}{L} \right)^{y_i} - \left( \tfrac{\eps}{L} \right)^{y_j + y_k}
      ] & \text{if $y_i \ne y_j + y_k$}\\
      C_{jk}{}^i \left( \tfrac{\eps}{L} \right)^{y_j + y_k} \ln{ \left( \tfrac{\eps}{L} \right)} & \text{if $y_i = y_j + y_k$}
   \end{cases}
  \; .
\end{align}
Hence,
\begin{align}
   \left( \tfrac{\eps}{L} \right)^{-y_i} \lam ^b_i =  \lam _i(L) +& \sum_{\tiny{\begin{array}{c} jk \\  y_i \ne y_j + y_k \end{array}}} \left( \frac{1}{y_i -
  y_j - y_k}C_{jk}{}^i ( 1 -  \left( \tfrac{\eps}{L} \right)^{-y_i +y_j + y_k} ) \right) \lam _j(L)
  \lam _k(L) \notag \\
  +& \sum_{\tiny{\begin{array}{c} jk \\  y_i = y_j + y_k \end{array}}}  C_{jk}{}^i \ln{\left( \tfrac{\eps}{L} \right)} \lam _j(L)
  \lam _k(L) \; .
\end{align}

\subsection{Links and Relevant Operators}
\label{sec:lro}

In this section we show that the number of links in a diagram of Behrend and Pearce is equal to the number of relevant operators on the appropriate conformal boundary condition.  Not only is this true for the A-series mentioned in the text, but also for the D and E models too. 

For the details of statements made in this section, we refer the reader to \cite{Beh99} and \cite{Beh00}. \\

In the notation of \cite{Beh99}, we look at the model $M(A_{m-1},G)$ with $G$ an ADE graph with Coexter number $g=m+1$.  For the A-Series $G=A_m$.  Note that we will not consider the unitary models with $g=m-1$ because as of this writing, they lack a lattice realisation with suitable boundary weights.

Let $G_{ab}$ be the adjacency matrix of $G$ which defines the fused adjacency matrix $F_{ab}^r$,
\begin{align}
  F_{ab}^{r+1} =  \sum_c G_{ac} F_{cb}^r - F_{ab}^{r-1} \; , 
  \qquad F_{ab}^1=\One \; , 
  \qquad F_{ab}^2=G_{ab} \; ,
  \label{eq:defnF}
\end{align}
The $F^r$ are symmetric matrices with positive integer entries and $F^{m+1}=0$.
In the case $G=A_m$, $F_{ab}^{r}$ is given by equation \eqref{eq:defnFA}. 

The ADE graph $G$ is bicolourable; that is to say, one may assign to each node $a \in G$, a parity $\pi_a \in \{-1,1\}$ such that adjacent nodes have opposite parity,
$G_{ab} > 0 \implies \pi_a \pi_b = -1$.  This extends to $F^r$:
\begin{align}
  F_{ab}^r > 0 \implies  \pi_a \pi_b = (-1)^{r+1} \; .
  \label{eq:bicolour}
\end{align}

As matrices, the $F^r$ form a representation of the fusion algebra,
\begin{align}
  F^r F^s = \sum_t N(m+1)_{rs}{}^t F^t \; , \label{eq:repfuse}
\end{align}
where $N(m)_{rs}{}^t$ is given in \eqref{eq:bign}.  One may also write the boundary spectrum of a general minimal model in terms of these matrices \cite{Beh99} (c.f. \eqref{eq:littlen}),
\begin{align}
   n_{(r,s) (r',s')}{}^{(r'',s'')}= N(m)_{r,r'}{}^{r''} F_{s's''}^{s} + N(m)_{m-r,r'}{}^{r''} F_{s's''}^{m+1-s} \; ,
  \label{eq:spec}
\end{align} \\

Following \cite{Beh00}, we label the nodes of $G$ by $1,2,\ldots,n$ then the subgraph
associated to the Cardy boundary condition $(r,s)$ has the adjacency matrix,
\begin{align}
  B(r,s)_{ab} =  F_{sa}^r G_{ab} F_{bs}^{r+1} \; .
\end{align} \\

We now turn to our theorem.  From \eqref{eq:h}, relevant fields in a boundary minimal model have the form $\phi_{r,r+t}$ for $t \in \{ -1,0,1,2 \}$.  Using \eqref{eq:spec} we find the number of relevant fields is given by,
\begin{align}
  \text{\# rel. fields on } (r,s) = \sum_{\tiny{\begin{array}{c} k \\ t \in  \{ -1,0,1,2 \} \end{array}}}  n_{(k,k+t) (r,s)}{}^{(r,s)}
  = \sum_k N(m)_{k r}{}^{r} ( F_{ss}^k + F_{ss}^{k+2} ) \; ,
  \label{eq:nrel}
\end{align} 
where the sum over $k$ in the first expression is chosen as to avoid over counting due to the identification $(r,s) \leftrightarrow (m-r,m+1-s)$.  In obtaining the second expression we have used the bicolourability of $G$. \\

On the other hand, counting the number of links in this subgraph with multiplicities we find,
\begin{align}
  \text{\# links} = \sum_{a,b} F_{sa}^r G_{ab} F_{bs}^{r+1} \; .
  \label{eq:nlinks}
\end{align}
As an aside, there is no over counting in \eqref{eq:nlinks} because the bicolourability of $G$ implies that $(a,b) \in B \implies (b,a) \notin B$.  From \eqref{eq:nlinks}, \eqref{eq:defnF} and \eqref{eq:repfuse}, 
\begin{align}
  \text{\# links} &=  \sum_{a} F_{sa}^r (F_{as}^{r} + F_{as}^{r+2}) 
  = \sum_k  N(m+1)_{rr}{}^k F_{ss}^k + N(m+1)_{r,r+2}{}^k F_{ss}^k
  \label{eq:nlinks2} \; .
\end{align}
To make the connection with \eqref{eq:nrel}, we consider each term in turn.  Using the explicit realisation of $N(m)$, \eqref{eq:bign}, we write,
\begin{align}
  \sum_k  N(m+1)_{rr}{}^k F_{ss}^k &
  = \sum_k  N(m)_{rr}{}^k F_{ss}^k + Q_1 \; , \\ 
 Q_1 &= \begin{cases} F_{ss}^{2(m-r)+1} \qquad \text{when} \qquad r \ge \tfrac 12 (m+1) \; , \\
   0 \qquad \qquad \qquad \text{otherwise} \; . \end{cases} 
\end{align}
and,
\begin{align}
  \sum_k  N(m+1)_{r,r+2}{}^k F_{ss}^k &
  = \sum_k  N(m)_{r,r}{}^k F_{ss}^{k+2} - Q_2 \; , \\
 Q_2 &= \begin{cases} F_{ss}^{2(m-r)+1} \qquad \text{when} \qquad r \ge \tfrac 12 m \; , \\
   0 \qquad \qquad \qquad \text{otherwise} \; . \end{cases}
\end{align} 
Now, $Q_1$ and $Q_2$ cancel except when $r=\tfrac 12 m$, $Q_2=F^{m+1}=0$.   Putting this all together,
\begin{align}
  \text{\# links} = \sum_k  N(m)_{rr}{}^k ( F_{ss}^k + F_{ss}^{k+2} ) \; ,
\end{align}
which is equal to \eqref{eq:nrel} by the cyclic nature of $N(m)$. \\

\end{document}